\documentclass[
 reprint,
 amsmath,amssymb,
 aps,
prb,
floatfix,longbibliography
]{revtex4-2}
\usepackage{amsmath}
\usepackage{amsfonts}
\usepackage{graphicx}
\usepackage{appendix}
\usepackage{dsfont}
\usepackage{amssymb}
\usepackage{bm}
\usepackage{xcolor}
\usepackage{ulem}
\usepackage{soul}
\usepackage{mathrsfs}
\usepackage{amsmath}
\usepackage{esint}
\usepackage{natbib}
\usepackage{nccmath}
\usepackage{array}
\newcommand{\beq}{\begin{equation}}
\newcommand{\eneq}{\end{equation}}
\newcommand{\be}{\begin{equation}}
\newcommand{\ee}{\end{equation}}
\newcommand{\bea}{\begin{eqnarray}}
\newcommand{\eea}{\end{eqnarray}}

\usepackage{multirow}
\usepackage{wasysym}

\begin{document}

\title{Spin and thermal current scaling at a $Y$-junction of XX spin chains}
 
\author{Domenico Giuliano}
\affiliation{Institut f\"ur Theoretische Physik, Heinrich-Heine-Universit\"at, 40225 D\"usseldorf, Germany}
\affiliation{Dipartimento di Fisica, Universit\`a della Calabria Arcavacata di 
Rende I-87036, Cosenza, Italy}
\affiliation{
I.N.F.N., Gruppo collegato di Cosenza 
Arcavacata di Rende I-87036, Cosenza, Italy  }
\author{Francesco Buccheri}
\affiliation{
Dipartimento di Scienza Applicata e Tecnologia, Politecnico di Torino,
Corso Duca degli Abruzzi 24, 10129, Torino, Italy}
\affiliation{INFN Sezione di Torino, Via P. Giuria 1, I-10125, Torino, Italy}
\begin{abstract}
We study the boundary phase diagram and the low-temperature heat and magnetization transport at a $Y$-junction of XX spin chains.
 Depending on the magnetization axis anisotropy  between the magnetic exchange interactions at the junction, the system exhibits two different strong-coupling regimes at low energies/temperatures,  hosting a remarkable realization of, respectively,  the  overscreened (topological)  four- and to the two-channel Kondo fixed points,
 both robust against anisotropies in the chain parameters and/or in the boundary couplings. Using renormalization group arguments combined with
  boundary conformal field theory methods, we  show the instability of the former under any  XY-type anisotropy at the junction. 
 Our system allows for using   the low-temperature spin and the heat conductances to evidence the fractionalization of the elementary 
  excitations at the four-channel Kondo fixed point by means of the magnetic Wiedemann-Franz law. 
We caution that the instability under XY anisotropy may hinder the detection of the phenomenology related to the four-channel 
Kondo effect, therefore requiring careful control in experimental realizations.
\end{abstract}
\maketitle

\section{Introduction}
\label{intro}

Energy flows in quantum systems are an important object of study in modern condensed matter physics: indeed, they lie at the heart of fundamental questions of out-of-equilibrium statistical physics. 
Of special interest for practical applications are one-dimensional models \cite{Gallavotti2007}. For example, driven spin chains can act as heat engines \cite{Barra2015}, which in theory allow for the conversion between energy and work at the quantum level \cite{Campisi2016,Arrachea2023}. Anomalous heat conductance can be present in one dimension, in the presence of momentum-conserving lattices \cite{Narayan2002}. In addition, heat currents 
reflect the conservation laws of  the systems at hand \cite{Zotos1997,Shimshoni2003,Prosen2011}.
Many platforms support one-dimensional heat conduction channels. Realizations range from carbon nanotubes \cite{Bockrath1999}, atomic chains \cite{Blumenstein2011}, metallic quantum wires, conventional and topological \cite{vanWees1988,Demartino2021}, as well as one-dimensional edge channels in graphene \cite{Wakabayashi2009}.

In this study, we will be concerned, in particular, with spin chain materials and their cold-atomic emulators. 
The investigation of heat transport in magnetic systems has indeed elicited strong interest after the discovery of the spin-Seebeck effect \cite{Uchida2008}, and established the whole field of spin caloritronics \cite{Bauer2012}. Remarkably, heat and magnetization currents 
have been demonstrated to be controllable at the microscopic level by magnetic fields and dissipation \cite{An2013,Chen2021,Zou2024}. 
Even in purely spintronics applications, thermal management is an important aspect related to energy efficiency and performance \cite{Uchida2021}. Recent studies on caloric effects in spin chains suggest the potential of this class of systems to control heat 
fluxes at the quantum level \cite{Konar2022,Zvyagin2024}.
Such a control is also relevant for the new possibilities that atomic physics offers, in connection to the quantum technologies 
based on optically trapped ultracold atoms \cite{Zhang2022,Bu2024}. Due to the extremely low temperatures involved, all such 
platforms require careful and precise control of the energy flows within the system \cite{Loft2016,Marchukov2016}.
\begin{figure}
    \centering    \includegraphics[width=0.9\linewidth]{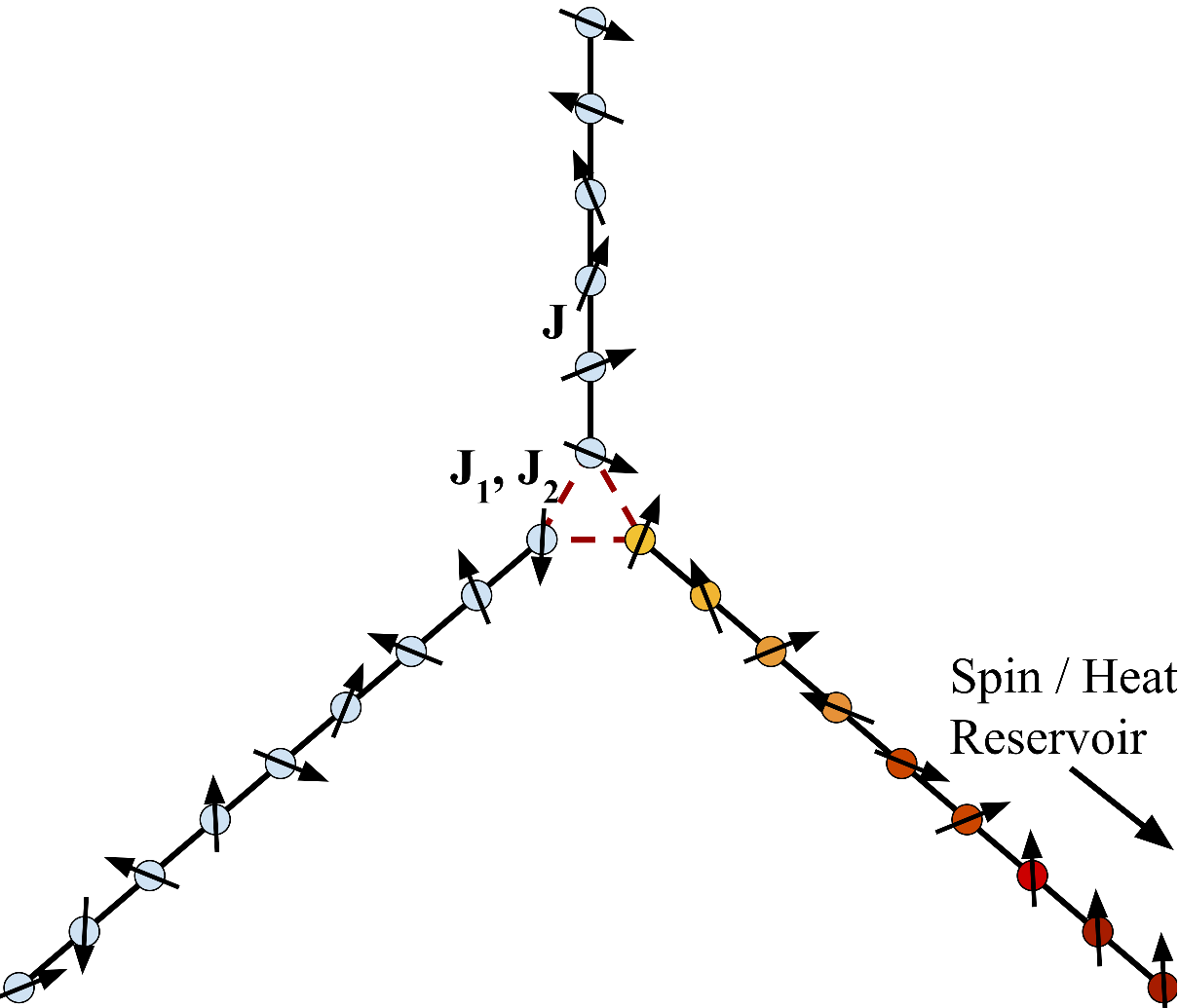}
    \caption{Sketch of a Y junction of semi-infinite spin chains. The magnetic spin exchange interaction $J$ along the three chains is the same along the $x$- and the $y$-directions and is represented by a black bond. In the central triangle the magnetic  exchange interaction can be different in the different representation, which we represent by dashed red lines. Each chain is connected to a magnetization and temperature reservoir.}
    \label{device}
\end{figure}

  The remarkable possibility of engineering networks of spin chains with nontrivial topological properties paves the way to the potential
implementation of simple structures, such as Y junctions of chains, as building blocks of thermal quantum circuits with controllable properties. On the complementary 
side, the thermal and magnetic transport properties of the junction are strictly related to the topology of the junction itself, specifically to
the low-energy/temperature behavior of the system. Therefore,  synoptically mapping out the thermal and magnetic transport pattern across the system at various energy/temperature scales
provides an effective mean to map out the phase diagram of  the junction, which is in general not simple in insulating systems, such as quantum spin chains. 
Based on the above motivations,  
in this work, we analyze in detail the simplest
nontrivial magnetic circuit element, namely, the Y junction \cite{Oshikawa2006}, see Fig. \ref{device}. The problem of heat transport in junctions and graphs has been tackled from a classical perspective \cite{Horvat2012,Barra2001}, as well as from a field-theoretical one \cite{Bellazzini2008,Bellazzini2006,Buccheri2022,Giuliano2022}. The energy flows within the junction, arising from contact with reservoirs at different temperatures, can be described in terms of non-equilibrium steady states \cite{Mintchev2015,Mintchev2011}, in which the chiral low-energy excitations are in equilibrium with their sources. 
What motivates interest in spin chain junctions is that a variety of fixed points (FPs) are known, enabling control of magnetic excitation \cite{Buccheri2018,Buccheri2019}, as well as the creation of highly entangled states \cite{Crampe2013,Tsvelik2014}. In presence of a gapless spectrum, a full Kondo cloud can indeed develop and modify the conduction properties of the FP   \cite{Giuliano2013,Giuliano2020,Laflorencie2008,Gaines2025}.
This becomes manifest in a non-trivial thermal conductance, which remains nonetheless linear in temperature \cite{Giuliano2022}.

Insulating magnets can support ballistic magnonic heat and spin transport with universal behavior analogous to that of electrons \cite{Nakata2015,Nakata2017}. This is encoded in the magnonic Lorenz ratio
\begin{equation}\label{LorenzRatio}
\mathcal{L}_m = \frac{\mathcal{J}_Q}{T\,\mathcal{J}_S} \;\; ,
\end{equation}
\noindent
where $\mathcal{J}_Q$ is the heat current, $\mathcal{J}_S$ the spin current, and $T$ the absolute temperature. In the low-temperature limit, the ratio asymptotically approaches a universal constant: $\mathcal{L}_m = \frac{\pi^2}{3} \left( \frac{k_B}{g \mu_B} \right)^2$, 
where $g$ is the Land\'e $g$-factor and $\mu_B$ is the Bohr magneton. This result is analogous to the Wiedemann-Franz law (WF) for electrons. In the magnonic case, the Lorenz ratio is independent of microscopic details such as the magnon dispersion or exchange coupling, and depends only on fundamental constants. This indicates a form of universality in magnon transport across insulating ferromagnetic junctions and provides theoretical foundation for designing spin conductors and energy‑efficient spin caloritronic devices.
 
A key point in engineering efficient quantum devices is to make their behavior robust against fluctuations in microscopic system parameters, minor details on the system design, and so on. An effective way to achieve this goal is to work at an interacting renormalization group (RG) FP of 
its phase diagram. By means of a combined use of perturbative RG methods and conformal field theory techniques, several interacting FPs have been identified in the phase diagram of Y-junctions of quantum wires, spin chains and/or one-dimensional Josephson junction arrays. These exhibit several non-conventional properties, such as spin or charge fractionalization \cite{Chamon2003,Oshikawa2006,Hou2008,Cirillo2011}, 
or pairing of Cooper pairs \cite{Giuliano2009EPL,Giuliano2010}. Moreover, they can host nontrivial realizations of the 
multi-channel Kondo effect, including its topological version 
\cite{Tsvelik2013,Tsvelik2014,Tsvelik2014_2,Altland2013,Altland2014,Giuliano2013,Giuliano2020,Giuliano2020b,Guerci2021,Buccheri2022}. 
 When aiming at realizing the notrivial Kondo-like FP, it is essential to have control of the parameters at the junctions, due to the (marginal) relevance of some junction interaction. In general, the RG flow of the system toward a given attractive FP may strongly depend on the initial conditions on the system parameters. 
Nevertheless, when a (topological)  multichannel Kondo system is realized at a junction of quantum spin chains, it is known that the 
flow of the system toward the Kondo
fixed point is not affected by fluctuations of the system parameters at leading order and, more importantly, it is not affected by differences in the exchange interactions 
within each XX chain, as well as between two different chains \cite{Tsvelik2013,Altland2013,Buccheri2016,Beri2012}.  
However, the spin Y junction contain another type of anisotropy, namely, exchange anisotropy, which can arise from purely geometric factors \cite{Berkowitz1999}.

 Motivated by the above observations, in this work, we map out the phase diagram and the FPs of a Y-junction of XX spin chains  by making
a combined use of perturbative RG methods combined with conformal field theory methods. Doing so, we derive the spin
 and the heat conductance tensor at the FPs, with emphasis on the magnonic equivalent of the WF \cite{Nakata2015}. 
 Our results eventually enable us to discuss   in detail the physical properties of the various strongly interacting FPs of the system
 and, in particular, to  highlight the instability of the topological Kondo (TK) FP under XY interaction anisotropy   at the junction, 
 which prompts special care in tuning the microscopic parameters in experiments.
 
 The paper is organized as follows:
 in Section \ref{modham}, we introduce the model Hamiltonian of our system, as well as the spin and heat currents. In Section \ref{ph_dia}, we present the phase diagram of the junction. The following Sections contain the heat conductance and the discussion of the WF around the
 various FPs. In particular, in Section \ref{pefu} we lay out the perturbative approach around the disconnected FP, while in Section
  \ref{s_fixed} and  \ref{sicatopo} we respectively discuss the Ising and the TK FPs. We summarize our results in Section 
  \ref{concl}. Various Appendixes contain further details about the Keldysh formalism, the RG procedure and
   the effective theories at the FPs.

\section{Model of the junction}
\label{modham}

We sketch our system in Fig.\ref{device}. The junction is made out of three XX chains connected to each other in a $Y$-shaped
junction at one of their endpoints. The system Hamiltonian is therefore composed of 
a free chain term, which only describes the three chains, 
$H_{\rm XX3} = \sum_{\lambda = 1}^3 H_{\rm XX,\lambda}$, 
with $\lambda$ being the chain index, and of a boundary term $H_K$, describing the 
junction between the three chains. The relevant parameters entering each chain Hamiltonian are 
the exchange interaction strength $J$ between nearest neighboring sites (that is planar
in the spin space), and the applied magnetic field $H$ (that lies along the $z$ axis in 
spin space). As a result, the ``bulk'' Hamiltonian for a single chain is given by 
\beq
H_{\rm XX,\lambda} =-J \sum_{j = 1}^{\ell -1} \{ S_{j,\lambda}^x S_{j+1,\lambda}^x + S_{j,\lambda}^y S_{j+1,\lambda}^y\} 
+H \sum_{j =1}^\ell S_{j,\lambda}^z \;  , 
\label{mh.1}
\eneq
\noindent
with $J$ ($>0$) being the magnetic exchange coupling strength and $H$ the applied Zeeman field. 
In Eq.(\ref{mh.1}) the quantum spin-1/2 operators $S_{j,\lambda}^a$ obey the commutation 
relations $[S_{j,\lambda}^a , S_{j',\lambda'}^{b} ] = i \sum_{c} \epsilon^{a,b,c} \delta_{j,j'} \delta_{\lambda,\lambda'}
S_{\lambda,j}^{c}$, with  $\epsilon^{a,b,c}$ being the fully antisymmetric tensor. For simplicity,  $H$ is rescaled by the factor
 $g\mu_B/\hbar$ and is therefore dimensionally a frequency. With this definition, the Wiedemann-Franz law takes the form
 
\begin{equation}\label{WF}
\mathcal{L}_m = \frac{\pi^2}{3} \left( \frac{k_B}{\hbar} \right)^2 \:\: . 
\end{equation}
\noindent
Arrays of confined polar molecules interacting via dipolar spin-exchange provide an excellent way of simulating the spin Hamiltonian \eqref{mh.1} 
\cite{Micheli2006,Gorshkov2011,Yan2013}. One-dimensional arrays of ion microtraps can approximately realize \eqref{mh.1} as well, provided the transverse trapping frequencies are sufficiently large \cite{Porras2004}. Finally, we mention that spin-up and -down states can also be encoded in two atomic states of a Rydberg atom, and dipole-dipole interaction were shown to be approximately described by the Hamiltonian \eqref{mh.1} \cite{Barredo2015}.
We also mention that the capacitive coupling between qubits tuned in resonance with each other generates the exchange interaction in \eqref{mh.1} \cite{Blais2021}.
Also, here we do not account for possible fluctuations in the exchange interactions within each XX chain, as well as between two different chains, as they are irrelevant in RG sense \cite{Tsvelik2013,Altland2013,Buccheri2016,Beri2012}.

As for the boundary Hamiltonian $H_K$ describing the junction, in full generality we allow for an asymmetry between 
the $x$- and the $y$-axis in spin space, that is, we set:

\beq
H_K = - J_1 \sum_{\lambda=1}^3 S_{\lambda,1}^x S_{\lambda +1,1}^x - J_2 \sum_{\lambda=1}^3 S_{\lambda,1}^y
S_{\lambda+1,1}^y 
\;\; , 
\label{mh.2}
\eneq
\noindent
with $\lambda + 3 \equiv \lambda$ understood when summing over the chain index.  
Trapped ions \cite{Micheli2006} and polar molecules \cite{Kotibhaskar2024} can simulate also anisotropic XY-type interactions when applying microwave near the resonance between vibrational modes or exploiting Raman transitions.
Finally, in Rydberg atom realizations, the XY coupling anisotropy can be engineered via microwave Floquet drive pulses \cite{Nishad2023}.
 
To study the transport properties of our system, it is
useful to trade it for a junction of lattice fermionic chains, possibly with a nontrivial boundary interaction
term arising when describing the system in fermionic coordinates \cite{Tsvelik2013,Giuliano2013,Tsvelik2014,Tsvelik2014_2,Giuliano2016,Giuliano2016_2,Buccheri2019} . 
In order to do so,  one has  to extend the Jordan-Winger (JW) transformation to a junction of three quantum spin  chains by introducing three additional Klein factors, 
$\{\eta_1,\eta_2,\eta_3\}$ \cite{Crampe2013}, so that the transformations from spin to fermionic variables read

\begin{eqnarray}
S_{j,\lambda}^+ &=& i \eta_\lambda c_{\lambda ,j}^\dagger e^{i\pi \sum_{r=1}^{j-1} c_{\lambda,r}^\dagger c_{\lambda , r}} \nonumber \\
S_{j,\lambda}^- &=& i \eta_\lambda c_{\lambda ,j} e^{i\pi \sum_{r=1}^{j-1} c_{\lambda,r}^\dagger c_{\lambda , r}} \nonumber \\
S_{j,\lambda}^z &=& c_{\lambda,j}^\dagger c_{\lambda,j}-\frac{1}{2} 
\:\: ,
\label{mh.3}
\end{eqnarray}
\noindent
with $S_{j,\lambda}^\pm = S_{j,\lambda}^x \pm i S_{j,\lambda}^y$. The
JW fermions satisfy canonical anticommutation relations 
\beq
\{c_{j,\lambda},c_{j',\lambda'}^\dagger \}=\delta_{j,j'}\delta_{\lambda,\lambda'}
\;\; , 
\label{mh.4}
\eneq
\noindent
while the Klein factors satify the algebra
\beq
\{\eta_\lambda , \eta_{\lambda'} \} = 2 \delta_{\lambda,\lambda'} 
\;\; , 
\label{mh.5}
\eneq
\noindent
Eqs.(\ref{mh.4},\ref{mh.5}) imply that  the canonical commutation relations for the quantum spin operators 
are naturally satisfied. When implementing Eqs.(\ref{mh.3}) the Klein factors disappear from the bulk 
Hamiltonian for the three chains.
In JW fermionic coordinates we obtain 

\begin{eqnarray}
H_{\rm XX,\lambda,f} &=& \sum_{\lambda=1}^3 \{ - J \sum_{j=1}^{\ell-1} [c_{j,\lambda}^\dagger c_{j+1,\lambda} 
+ c_{j+1,\lambda}^\dagger c_{j,\lambda} ] \nonumber \\
&&+ H \sum_{j=1}^\ell c_{j,\lambda}^\dagger c_{j,\lambda} + {\rm const} \} 
\:\: ,
\label{mh.6}
\end{eqnarray}
\noindent
with the over-all constant contribution being unimportant for the following calculations and, therefore, 
neglected henceforth. At variance, the Klein factors do appear in $H_K$ which, in fermionic coordinates, 
reads

\begin{eqnarray}\label{mh.7}
H_{K,f} &=&\frac{1}{2} \: \sum_{\lambda =1}^3 \biggl\{ [-i\eta_\lambda \eta_{\lambda +1} ] \times \\
&& \biggl[J_1 [-i(c_{\lambda,1}^\dagger +c_{\lambda,1} )(c_{\lambda+1,1}^\dagger 
+ c_{\lambda+1,1} )]\nonumber \\
&&+ J_2  [i (c_{\lambda,1}^\dagger - c_{\lambda,1} )( c_{\lambda+1,1}^\dagger -
c_{\lambda+1,1} ) ]\biggr]\biggr\} 
\:\: . \nonumber
\end{eqnarray}
\noindent
In the explicit expression for $H_{K,f}$ in Eq.(\ref{mh.7}) we identify a 
Kondo-like  interaction, with the
``standard'' Kondo impurity spin replaced by the ``topological'' spin operator ${\cal T}_\lambda 
= - i \eta_{\lambda + 1 } \eta_{\lambda + 2} $ (with $\lambda + 3 \equiv \lambda$) 
\cite{Beri2012,Altland2014,Buccheri2012,Giuliano2013}.  The topological spin is  
coupled to two, independent spin operators made out of the lattice fermionic fields at 
the junction site, $j=1$. As a result, one infers  how the   Kondo interaction 
Hamiltonian $H_{K,f}$ corresponds to 
a doubled-channel version of the two-channel Kondo interaction, as proposed in 
\cite{Coleman1995}, that is, to a topological, four-channel Kondo model \cite{Crampe2013,Giuliano2016,Buccheri2016}.

As they are the key observable quantities we use throughout our paper to characterize the various
FPs, we now  derive the lattice formulas for the conserved spin- and heat-current operators. 
Let $j$ and $j+1$ denote sites in the bulk of a chain (that is, both different from 
$1$ and $\ell$), to site $j$ we associate the spin density in the $z$ direction given by 
$S_{j,\lambda}^z = c_{j,\lambda}^\dagger c_{j,\lambda} - \frac{1}{2}$. At the same time, to the link 
$(j,j+1)$ we associate the energy density given by 

\begin{eqnarray}
{\cal E}_{(j,j+1),\lambda} &=& - J \{c_{j,\lambda}^\dagger c_{j+1,\lambda} 
+ c_{j+1,\lambda}^\dagger c_{j,\lambda} \} \nonumber \\
&+& \frac{H}{2} \{c_{j,\lambda}^\dagger c_{j,\lambda} + c_{j+1,\lambda}^\dagger 
c_{j+1,\lambda} \} 
\:\: . 
\label{mh.8}
\end{eqnarray}
\noindent
If $1<j<\ell$, the Heisenberg equation of motion for the operator $c_{j,\lambda} (t)$ is given by 

\beq
i \hbar \frac{\partial c_{j,\lambda} (t)}{\partial t} = 
- J \{ c_{j+1,\lambda} (t) + c_{j-1,\lambda} (t) \} + H c_{j,\lambda} (t ) 
\:\: . 
\label{mh.9}
\eneq
\noindent
From Eqs.(\ref{mh.9}) we readily recover the continuity equations for the spin- and for the energy-density 
operators, that is

\beq
\frac{d S_{j,\lambda}^z}{d t } = - \{ {\cal J}_{S,j+1,\lambda} - {\cal J}_{S,j,\lambda} \} 
\;\; , 
\label{mh.10}
\eneq
\noindent
for the spin density, and 

\beq
\frac{d {\cal E}_{(j,j+1),\lambda}}{dt} = - \{ {\cal J}_{Q,(j+1,1),\lambda} - {\cal J}_{Q,(j,j-1),\lambda} \}
\;\; , 
\label{mh.11}
\eneq
\noindent
for the energy density, with 

\begin{eqnarray}
{\cal J}_{S,j,\lambda} &=& - i J \{c_{j,\lambda}^\dagger c_{j+1,\lambda} - c_{j+1,\lambda}^\dagger c_{j,\lambda} \} \nonumber \\
{\cal J}_{Q,(j+1,j),\lambda} &=& i \frac{J^2}{ \hbar } \{c_{j ,\lambda}^\dagger c_{j+2,\lambda} 
- c_{j+2,\lambda}^\dagger c_{j ,\lambda} \} 
\:\: . 
\label{mh.12}
\end{eqnarray}
\noindent
We now derive the phase diagram of our system, a task for which we have to refer to
the RG equations derived in Appendix \ref{RGK}.

\section{Phase diagram of the $Y$-junction of XX chains}
\label{ph_dia}

The simplest FP in the phase diagram of our junction 
(the ``disconnected FP, DFP)
 corresponds to the full disconnection 
of the three chains from each other, that is, to $J_1=J_2=0$. 
A nonzero $J_1$ and/or $J_2$ yield to nonperturbative effects which 
we account within  RG approach
in the corresponding running couplings. In Appendix \ref{gefu} and \ref{RGK} we
present in detail the various steps leading to the RG equations. As a result, we
find that, on lowering the high-energy cutoff from $D$ to $D-\delta D$, the coupling strengths are
renormalized according to the flow equations

\beq
\frac{d J_{1,2}}{d D} = -\frac{6 \varphi (D,H)}{\pi J D} J_{1,2}^2
\;\; , 
\label{cmh.9}
\eneq
\noindent
with 

\begin{eqnarray}\label{cmh.10}
\varphi (D,H) &=& \left\{1+\frac{D^2-H^2}{2 J^2} + \left[\frac{D^2+H^2}{4J^2}\right]^4 \right\}^\frac{1}{4} 
 \\
&&\times
\cos \left[\frac{1}{2} {\rm atan} \left(\frac{2HD}{4J^2+D2-H^2} \right) \right]
\;\; . \nonumber
\end{eqnarray}
\noindent
Eqs.(\ref{cmh.9}) imply a growth of the running couplings on lowering the cutoff scale $D$. 
To second-order in the boundary couplings themselves, the equations decouple. The Kondo 
temperature pertinent to each channel, $T_{K,(1,2)}$   can be computed 
by using the implicit equation 

\beq
\frac{\pi J}{6 J_{1,2}} = \intop_{k_B T_{K,(1,2)} }^{D_0} \: \left[\frac{ \varphi (u,H)}{u} \right] \: d u \;\; , 
\label{cmh.11} 
\eneq
\noindent
with $D_0 \sim 2J$ (see Appendix \ref{RGK} for details) which implies that, the higher temperature corresponds
to the larger (``bare'') boundary coupling. 

Let us, now, assume, for the sake of the following discussion, that 
$J_1 (D_0 ) > J_2 (D_0)$. Integrating Eqs.(\ref{cmh.9}), we find that 
$J_1 (D)$ first crosses over to the strongly coupled regime at the corresponding
Kondo scale $D=D_{*,1} = k_B T_{K,1}$. As we discuss in detail in Appendix 
\ref{islike}, at scales $<D_{*,1}$ the junction flows towards the ``Ising-like'' FP ${\cal I}_1$, in which half of the real fermionic modes
out of which the complex lead fermions are formed, conspire to screen the topological impurity spin, similarly with what happens
at a $Y$ junction of three Ising chains \cite{Tsvelik2013,Tsvelik2014_2,Giuliano2013}, while the other half contribute the over-all system dynamics only perturbatively in  $J_2$. 
At variance, the complementary situation, in which $J_2 (D_0) > J_1 (D_0)$, implies a flow of the junction 
toward    the complementary  ${\cal I}_2$ Ising-like FP. 

\begin{figure}[h]
    \centering    \includegraphics[width=0.75\linewidth]{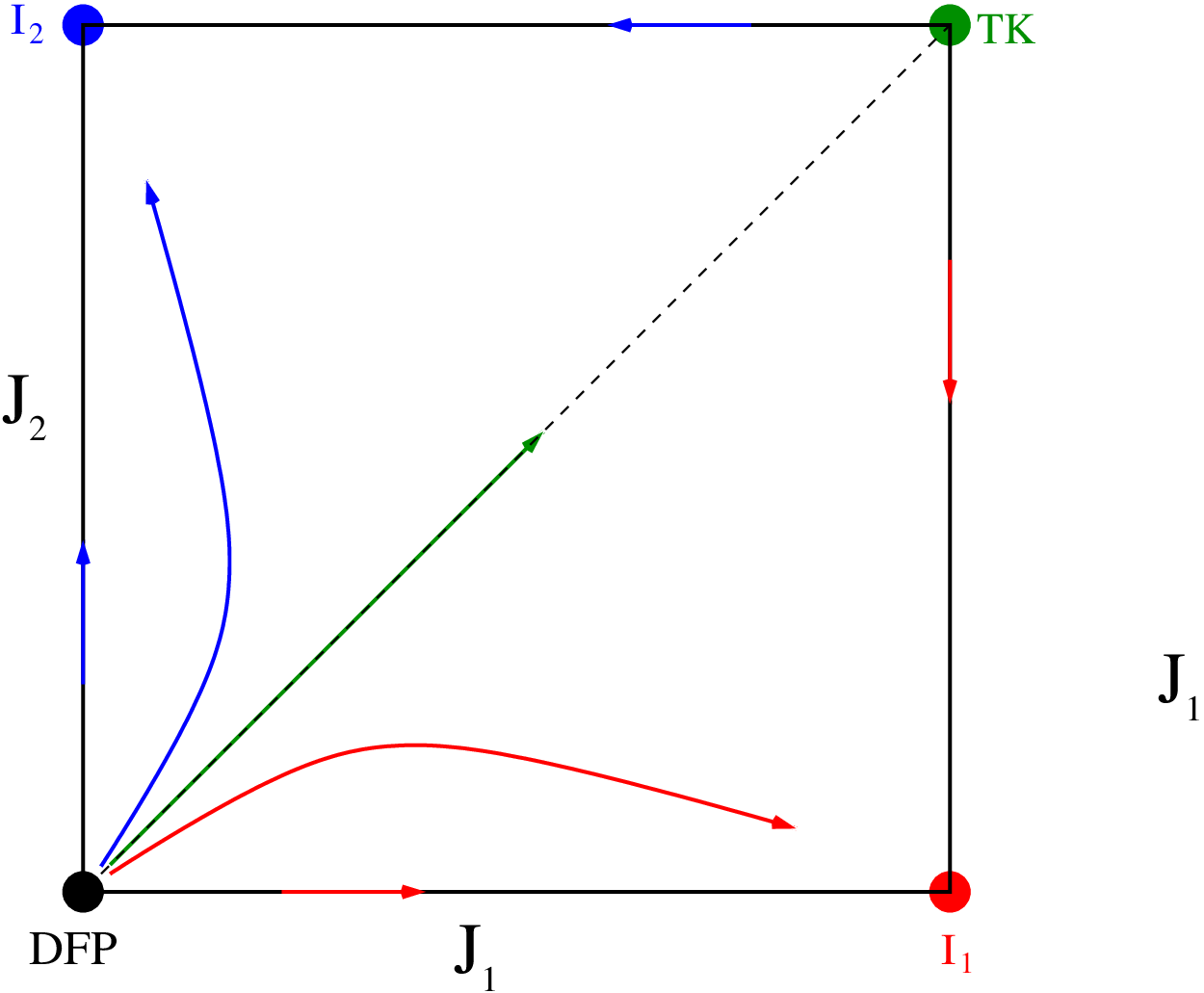}
    \caption{Sketch of the RG trajectories and of the 
    phase diagram of the junction in the $J_1$-$J_2$ plane. There are four FPs: the disconnected FP,
    drawn as a full black dot, at $J_1=J_2=0$, which is fully unstable against the Kondo boundary interaction, 
    the TK FP, drawn as a full green  dot, which is reached along the RG trajectories 
    at $J_1=J_2$, but is unsable against any asymmetry between couplings $J_1$ and $J_2$, the two Ising-like FPs 
    ${\cal I}_1$ and ${\cal I}_2$, respectively drawn as a full blue and red dot, which are stable and are reached 
    along the RG flow from whatever point in parameter space with $J_1>J_2$, or $J_1<J_2$, respectively. }
    \label{pha_dia}
\end{figure}

Finally, as we discuss  in Appendix \ref{ren_sym}, when $J_1 (D_0)= J_2 (D_0)$  the junction
flows toward a FP analogous to the one corresponding to the strongly coupled regime of the 
TK model (TKM)
\cite{Beri2012,Altland2013,Buccheri2022}.  At the TKM, the hybridization between the topological spin, determined by the Klein factors, and 
the lead degrees of freedom yields a crossover in the boundary conditions of the  ``relative'' bosonic fields 
$\phi_\lambda (x) - \phi_{\lambda +1}(x)$ at $x=0$ from Neumann boundary conditions  
(corresponding to the DFP) to Dirichlet boundary conditions (corresponding to the strongly coupled TKM 
FP).  Thus, on one hand we conclude that, as soon as one turns on nonzero $J_1$ and/or $J_2$, 
the DFP is unstable, on the other hand, that whether the system flows towards either ${\cal I}_1$, or ${\cal I}_2$, or 
to the TKM FP,  depends on the relative values of the boundary couplings at the reference scale.
Based on the perturbative results  we therefore infer  the phase diagram of Fig.\ref{pha_dia}, 
where we sketch the RG flow of the running coupling 
and highlight the various FPs that emerge from a minimal assumption on the topology of the phase diagram itself. 

\noindent
The phase diagram we draw in Fig.\ref{pha_dia} implies that, while both ${\cal I}_1$ and ${\cal I}_2$ are attractive FPs, 
there has to be a line, in parameter space, separating the RG trajectories pointing to either one of them. Apparently, 
the most natural choice is identifying the line with the diagonal $J_1=J_2$.  As we discuss in the following, 
this  implies that the whole line, 
including the strongly coupled TK FP, must be unstable against any nonzero $J_1-J_2$.  

To study the stability of the various FPs, we first derive all possible boundary 
operators allowed at a specific FP from pertinent   lattice model Hamiltonians   (corresponding to the various putative 
FPs) and, then, we employ 
the delayed evaluation of boundary conditions (DEBC) method, originally 
introduced in the context of junctions of quantum wires in Ref.\cite{Oshikawa2006}.  
To illustrate how our method works, let us begin with the DFP.  Formally, 
the disconnected junction  is described by  imposing open boundary conditions at the junction site 
  on all three the  $c_{j,\lambda}$, that is, $c_{0,\lambda} = c_{0,\lambda}^\dagger 
= 0$, $\forall \lambda = 1,,2,3$. Retaining only low-lying modes around the Fermi points, one recovers
the expansion of Eq.(\ref{ap.c.1}) of Appendix \ref{gefu}, that is  

\beq
c_{j,\lambda} \approx e^{ik_F j } \psi_{R,\lambda} (x_j ) + e^{-ik_F j} \psi_{L,\lambda} (x_j)  
\;\; , 
\label{cmh.12}
\eneq
\noindent 
with  the dynamics of the chiral fermionic fields   described by the Hamiltonian 

\begin{eqnarray}
H_0 &=&- i \hbar v \: \sum_{\lambda = 1}^3 \: \int_0^\ell \: d x \: \{\psi_{R,\lambda}^\dagger (x) \partial_x \psi_{R,\lambda} (x) 
\nonumber \\
&&- \psi_{L,\lambda}^\dagger (x) \partial_x \psi_{L,\lambda} (x) \} 
\:\: ,
\label{cmh.12bis}
\end{eqnarray}
\noindent
and $-2J\cos(k_F)+H=0$,  $x_j = aj$ (and the lattice step $a$ set to 1 henceforth). The open boundary conditions at 
the junction imply 

\begin{eqnarray}
&& \psi_{R,\lambda}(0) + \psi_{L,\lambda} (0) = 0 \nonumber \\
&& \psi_{R,\lambda}^\dagger (0) + \psi_{L,\lambda}^\dagger (0) = 0 
\;\; . 
\label{cmh.13}
\end{eqnarray}
\noindent
Eqs.(\ref{cmh.13}) allow for   ``unfolding''  the chiral fermionic fields, that is, for  
introducing a triple of chiral fields $\{\psi_\lambda \}$, defined for $-\ell \leq x \leq x$,  
such that 

\beq
\psi_\lambda (x) = \theta (x) \psi_{R,\lambda} (x) - \theta (-x) \psi_{L,\lambda} (-x) \; , 
(-\ell \leq x \leq \ell )
\:\: . 
\label{cmh.14}
\eneq
\noindent
 Turning on a finite $J_1$ and/or $J_2$ and taking into account  Eqs.(\ref{cmh.12},\ref{cmh.13},\ref{cmh.14}),
  we obtain the corresponding boundary interaction Hamiltonian in the form 

\begin{eqnarray}
&& H_{K,f}  =  \sum_{\lambda = 1}^3 \: \Biggl\{  [-i \eta_\lambda \eta_{\lambda + 1} ]  \times 
\label{cmh.15}\\
&&\Biggl[  \left( \frac{\bar{J}_1 + \bar{J}_2}{2} \right) \ [-i (\psi_\lambda^\dagger (0) \psi_{\lambda +1}(0) 
- \psi_{\lambda +1}^\dagger (0) \psi_\lambda (0) ) ]  \nonumber \\
&+&  \left(\frac{\bar{J}_1 - \bar{J}_2}{2} \right) \: \sum_{\lambda = 1}^3 [ - i (\psi_\lambda (0) \psi_{\lambda + 1}(0) - 
\psi_{\lambda +1}^\dagger (0) \psi_\lambda^\dagger (0) ) ]\Biggr]\Biggr\} \nonumber 
\:\: ,
\end{eqnarray}
\noindent
with $\bar{J}_i = 4 \sin^2 (k_F) J_i$. The operator at the right-hand side of Eq.(\ref{cmh.15}) is marginally relevant, as shown by 
our  RG analysis of Appendix \ref{RGK}, which implies a corresponding flow away from 
the DFP whenever either $J_1$, or $J_2$ (or both) are $\neq 0$. This is enough to conclude that
 the DFP is a repulsive FP. 

We now consider either one of the equivalent, Ising-like FP, say ${\cal I}_1$. As we discuss in detail in  
Appendix \ref{islike},  ${\cal I}_1$ can be described in terms of 
the two triples of  unfolded (over the line $-\ell \leq x \leq \ell$) real fermionic fields 
$\bar{\xi}_\lambda (x),\bar{\zeta}_\lambda (x)$, defined in Eqs.(\ref{ap.z1},\ref{ap.z4}). 
 In terms of the unfolded fields, the bulk Hamiltonian and the spin and the thermal current operators are given by

\begin{eqnarray}
H_0 &=& - i \hbar v \: \sum_{\lambda =1}^3 \: \int_{-\ell}^\ell \: d x \: \{\bar{\xi}_\lambda (x) \partial_x \bar{\xi}_\lambda (x) + 
\bar{\zeta}_\lambda (x) \partial_x \bar{\zeta}_\lambda (x) \} \nonumber \\
{\cal J}^z_{S,\lambda} (x) &=& i\hbar v \: \sum_{a=\pm} \: \bar{\xi}_\lambda (ax) \bar{\zeta}_\lambda (ax)
\label{srois.16x}\\
{\cal J}_{Q,\lambda}(x) &=& -i\hbar v^2 \: \sum_{a=\pm 1} \: a \{\bar{\xi}_\lambda (ax) \partial_x \bar{\xi}_\lambda (ax) 
+  \bar{\zeta}_\lambda (ax) \partial_x \bar{\zeta}_\lambda (ax) \} \nonumber 
\;\; . 
\end{eqnarray}
\noindent
The leading boundary interaction at the ${\cal I}_1$ FP, which we derive in 
detail in Appendix \ref{islike}, takes the form 

\begin{eqnarray}\label{strois.17x}
H_{{\cal I}_1}  &=&  - {\cal A} [-i \prod_{\lambda = 1}^3 \bar{\xi}_\lambda (0) ]   {\cal Q} \nonumber \\
&&+ i {\cal B} \sum_{\lambda = 1}^3 \{ \bar{\xi}_\lambda (0) \bar{\zeta}_{\lambda +1} (0) \bar{\zeta}_{\lambda +2}(0) ] \}
{\cal Q}   
\;\; ,
\end{eqnarray}
\noindent
with ${\cal A},{\cal B}$ being numerical coefficients computed in Appendix \ref {islike} and 
${\cal Q}$ being a real, fermionic zero-mode operator. The operator in Eq.(\ref{strois.17}) 
has dimension  3/2, that is 
larger than the critical dimension (1) below which a boundary operator is relevant and, therefore, we conclude that ${\cal I}_1$ is a stable attractive FP of the phase diagram.  
The same conclusion is reached for the FP ${\cal I}_2$, once noting that it is mapped onto ${\cal I}_1$ by just
 swapping $\bar{\xi}_{\lambda} (x)$ and $\bar{\zeta}_{\lambda} (x)$ with each other. 
 Thus, as already remarked above, we see that the stability of both ${\cal I}_1$ and ${\cal I}_2$  necessarily implies 
 the existence of a   phase boundary in the
$J_1-J_2$ parameter space. \ Upon our ``maximally simplified'' assumption, by symmetry considerations, the natural location of the phase boundary is the line $J_1 = J_2$, which, in Fig.\ref{pha_dia}, is represented as a black, dashed line. 

In between the two stable phases there is the TK  FP, which we represent as a full, green dot, in 
Fig.\ref{pha_dia}. From the weak coupling regime, the TK FP is only accessible along RG trajectories originating 
from the  ``symmetric'' initial condition, $J_1 (D_0) = J_2 (D_0)$. To  describe it, it is useful to 
bosonize the lead Dirac fields, by introducing two triple of chiral, massless Klein-Gordon fields of opposite chirality, 
 $\varphi_{R,\lambda} (x)$ and $\varphi_{L,\lambda} (x)$, such that 
$\psi_{R,\lambda} ( x ) = \Gamma_\lambda e^{i \varphi_{R,\lambda} (x)}$, $\psi_{L,\lambda} (x) = \Gamma_\lambda
e^{i\varphi_{L,\lambda} (x)} $, with $\Gamma_\lambda$ being real-fermion Klein factors (see Appendix \ref{ren_sym} for details). 
 Within the bosonization picture the flow between the DFP and the TK FP corresponds to a 
 flow from Neumann boundary conditions in all three the channels, that is
 
 \beq
 \varphi_{R,\lambda} (0) = \varphi_{L,\lambda}(0) \;\; , \; \forall \lambda
 \;\; , 
 \label{cmh.20}
 \eneq
 \noindent
 at the DFP,  to (still)  Neumann boundary conditions in the center-of-mass channel, that is, 
 $\Phi_R (0) = \Phi_L (0)$, with $\Phi_{R(L)} (x) = \frac{1}{\sqrt{3}} \: \sum_{\lambda = 1}^2 \: \varphi_{R(L),\lambda} (x)$,
 and Dirichlet boundary conditions in the ``relative'' channels, $\chi_{R(L),1} (x) = \frac{1}{\sqrt{2}} \{
 \varphi_{R(L),1}(x) - \varphi_{R(L),2} (x) \}$, and $\chi_{R(L),2}(x) = \frac{1}{\sqrt{6}} \{\varphi_{R(L),1}(x) + \varphi_{R(L),2}(x) 
 - 2 \varphi_{R(L),3}(x) \}$, that is 
 
 \begin{eqnarray}
 &&\chi_{R,1} (0) + \chi_{L,1} (0) = 0 \nonumber \\
 && \chi_{R,2} (0) + \chi_{L,2} (0) = 0 
 \:\: . 
 \label{cmh.21}
 \end{eqnarray}
 \noindent
 A relevant boundary operator $\delta \tilde{H}_\Delta$ at the TK FP arises  as we show in 
 Appendix \ref{ren_sym}, when  $J_1 \neq J_2$ is broken. In this case, 
we obtain 
 
 \beq
 \delta \tilde{H}_\Delta = -4  \delta \bar{J} \: \cos \left[\frac{ \Phi_R (0)+\Phi_L (0)}{\sqrt{3}} \right]
 \:\: , 
 \label{cmh.22}
 \eneq
 \noindent
 with $\delta \bar{J} = \bar{J}_1 - \bar{J}_2$, with corresponding  scaling dimension $d_\delta J = \frac{2}{3} < 1$.  
 
 Putting together all the above results about the behavior of the system close to the various FP, 
 a ``minimal assumption'' yields the phase diagram of 
 Fig.\ref{pha_dia}. Specifically, we see that our $Y$-junction of XX chains can host 
 both the Kondo effect as realized at a junction of critical Ising chains \cite{Tsvelik2013}, as well as a remarkable spin
 version of the TK effect \cite{Buccheri2022}. The TK FP, here, is realized only along the symmetric line
 $J_1 = J_2$ and, in a sense, it works as a separating line between the two, stable Ising-like phases. 
 
 The peculiar topological structure of the phase diagram comes along with a peculiar behavior of 
 the spin- and of the heat -current pattern in the various phases, 
 which we discuss in detail in the following.

\section{Perturbative calculation of the spin- and of the heat-current pattern across the junction}
\label{pefu}

In junctions of quantum wires it has been by now well established how a synoptical analysis
of the charge- and of the heat-transport across the junction provides an effective mean to map out  
the phase diagram of the junction \cite{Buccheri2022,Giuliano2022}.   Here,  
we perform  a similar analysis for  our Y-junction of three XX spin chains, of course by
substituting the charge current  with the   $z$-component of the spin current. Importantly, 
while the spin current is conserved in the leads,  as we show below, 
allowing for different spin exchange strengths in the
$x$- and in the $y$-direction at the junction leads to phases in which spin conservation is violated,
such as the ones corresponding to the  ${\cal I}_1$ and the ${\cal I}_2$ FPs.

In this Section, we derive  the spin- and heat-current conductance within  linear response theory in
the applied magnetic fields and temperature biases, to leading order in $H_{K,f}$. We do so
by employing the Keldysh Green's function approach which is potentially amenable to extend  our derivation to nonequilbrium transport \cite{Wauters2023}.  
Eventually, we improve our perturbative result by 
substituting the ``bare'' interaction strengths with the running ones, obtained by solving 
Eqs.(\ref{cmh.9}). 

Retaining only low-energy, long-wavelength excitations around the Fermi point, we describe the
DFP in terms of the field operators in Eqs.(\ref{cmh.12}) and of the corresponding unfolded fields in
Eqs.(\ref{cmh.14}). Accordingly,   the spin- and 
the heat-current operators are given by

\begin{eqnarray}
{\cal J}^z_{S,\lambda} (x) &=&  \hbar  v \: \{\psi_\lambda^\dagger (x) \psi_\lambda (x) -
 \psi_\lambda^\dagger (-x) \psi_\lambda (-x) \} 
\nonumber 
\\ {\cal J}_{Q,\lambda} (x) &=& \frac{i \hbar v^2}{2} \: \{\psi_\lambda^\dagger (x) 
 \partial_x \psi_\lambda (x) - [\partial_x \psi_\lambda^\dagger (x) ]
\psi_\lambda (x) 
\label{fiet.4}\\
&& + \: \psi_\lambda^\dagger (-x)  \partial_x \psi_\lambda (-x) -
 [\partial_x \psi_\lambda^\dagger (-x) ]
\psi_\lambda (-x) \} \nonumber 
\:\: . 
\end{eqnarray}
\noindent
Denoting with ${\cal I}_{S,\lambda} (x)$ and ${\cal I}_{Q,\lambda} (x)$ the average values 
of ${\cal J}_{S,\lambda}^z (x)$ and of 
${\cal J}_{Q,\lambda} (x)$, respectively, we obtain:  
 
\begin{eqnarray}
I_{S,\lambda} (x) &=& -  \hbar v \:\int\: \frac{d \omega}{2 \pi} \:  \{ G^\lambda_{(-,+)} (x,x;\omega) - 
G^\lambda_{(-,+)} (-x,-x;\omega) \} \nonumber \\
I_{Q,\lambda} (x) &=& - i \hbar v^2 \: \int \: \frac{d \omega}{2 \pi} \: \{ \lim_{x'\to x} \partial_{x'} \: G^\lambda_{(-,+)} (x',x;\omega)   \nonumber \\
&&+
\lim_{x'\to -x} \partial_{x'} \: G^\lambda_{(-,+)} (x',-x;\omega) \}
\:\: ,
\label{fiet.9}
\end{eqnarray}
\noindent 
with the Keldysh-Green's functions in the mixed (real space - frequency)  representation defined as 

\beq
G^\lambda_{( \eta,\eta')} (x,x';\omega) = \int \: dt \: e^{i\omega t} \: G^\lambda_{(\eta,\eta')} ( x,x';t) 
\;\; ,
\label{fiet.x1}
\eneq
\noindent
with

\beq
G_{(\eta,\eta')}^\lambda (x,x';t) = \langle {\bf T}_K \psi_\lambda (x,t,\eta) \psi_\lambda^\dagger (x',0,\eta') 
e^{i \intop_{\cal K} \: d \tau \: H_{K,j} (\tau) } \rangle 
\;\; , 
\label{fiet.x2}
\eneq
\noindent
and $\eta,\eta'$ being Keldysh indices corresponding to the two branches over the Keldysh path ${\cal K}$,
over which ${\bf T}_K$ is the path-ordering operator (see
Appendix \ref{keld} for details). For $J_1 = J_2 = 0$, 
one gets $G^\lambda_{(\eta,\eta')} (x,x';\omega ) = g^\lambda_{(\eta,\eta')} (x,x';\omega)$, with the 
Green-Keldysh functions in the disconnected limit  listed in Eq.(\ref{fiet.5}) of Appendix \ref{gefu}. 
By direct inspection one finds that   the first nonzero contributions to the currents comes to second-order in
the $J_i$.  To 
 compute it, we need   the Keldysh-Green functions evaluated to the same order in the boundary
 couplings. These are given by

 \begin{eqnarray}
 && G^\lambda_{(\eta,\eta')} (x,x';\omega)  =  g^\lambda_{(\eta,\eta')} (x,x';\omega) \nonumber \\
&-& \frac{ (\bar{J}_1+\bar{J}_2)^2 }{4}  \: \sum_{\eta_1,\eta_2 = \pm 1} \: \eta_1 \eta_2 \: g^\lambda_{(\eta,\eta_1)} (x,0;\omega) 
g^\lambda_{(\eta_2,\eta')} (0,x';\omega) \nonumber \\
&\times&  \sum_{a = \pm 1} \: g^{\lambda +a}_{(\eta_1,\eta_2)} (0,0;\omega ) \nonumber \\
&+& \frac{ (-\bar{J}_1+\bar{J}_2)^2 }{2} 
\: \sum_{\eta_1,\eta_2 = \pm 1} \: \eta_1 \eta_2 \: g^\lambda_{(\eta,\eta_1)} (x,0;\omega) 
g^\lambda_{(\eta_2,\eta')} (0,x';\omega) \nonumber \\
&\times& \sum_{a = \pm 1} \: g^{\lambda +a}_{(\eta_2,\eta_1)} (0,0;-\omega ) 
\:\: . 
\label{fiet.8}
\end{eqnarray}
\noindent
To recover the spin- and the  heat-conductance, we compute the right-hand side of Eq.(\ref{fiet.8}) by assuming that 
the lead $\lambda$ is   biased by  an ``extra'' magnetic field $\delta H_\lambda$  and by an
extra temperature $\delta T_\lambda$  (with respect to the equilibrium values $H$ and $T$ that
are the same for all three the leads). Then,  
plugging the result of Eq.(\ref{fiet.8}) into Eqs.(\ref{fiet.9}) for the currents and expanding to linear order in
the $\delta H_\lambda$ and in the $\delta T_\lambda$, we obtain

\begin{eqnarray}
&&I_{S,\lambda} (x)  =  {\cal A}_S (T)  \{ (\bar{J}_1 + \bar{J}_2 )^2 [2 \delta H_\lambda - \delta H_{\lambda +1} - 
\delta H_{\lambda -1} ] \nonumber \\
&&+  (-\bar{J}_1 + \bar{J}_2)^2 [2 \delta H_\lambda + \delta H_{\lambda +1}+ 
\delta H_{\lambda -1} ]  \}
\;\; ,
\label{fiet.10}
\end{eqnarray}
\noindent
for the spin current, with 

\beq
{\cal A}_S (T) = \int \: \frac{ d \omega}{8 \pi \hbar } \: \left[ \frac{\beta }{v^2 \cosh^2 \left(\frac{\beta \omega}{2} \right)} \right] 
= \frac{1}{4\pi \hbar v^2} 
\:\: ,
\label{fiet.11}
\eneq
\noindent
and  $\bar{J}_i = 4 \sin^2 (k_F) J_i$, and 

\beq
I_{Q,\lambda} (x) = 2 (\bar{J}_1^2 + \bar{J}_2^2) \: {\cal A}_Q (T) 
\{ - 2 \delta T_\lambda + \delta T_{\lambda + 1} + \delta T_{\lambda -1} \} 
\;\; , 
\label{fiet.12}
\eneq
\noindent
with 

\beq
{\cal A}_{Q,\lambda} (T) = \frac{1}{4 k_B T^2 v^2 \hbar^3 } \: \int \: \frac{d \omega}{2 \pi} \: 
\left[\frac{\omega^2}{\cosh^2 \left(\frac{\beta \omega}{2} \right) } \right]
= \frac{  k_B^2 T \pi^2}{6 \pi \hbar^3 v^2} 
\:\: ,
\label{fiet.13}
\eneq
\noindent
for the heat  current.

\begin{figure}[h]
    \centering    \includegraphics[width=0.6\linewidth]{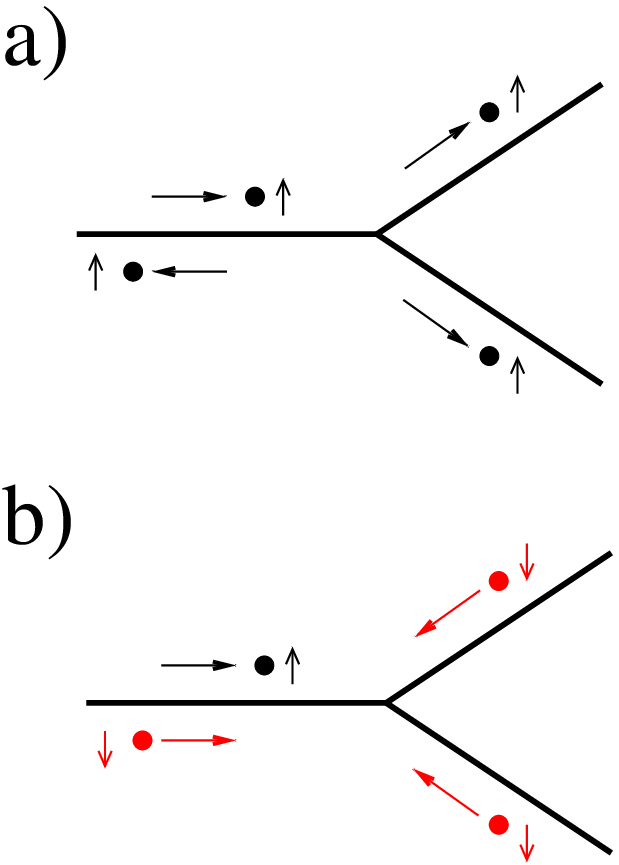}
    \caption{Sketch of the the elementary reflection/transmission processes supporting (perturbatively in the boundary 
    couplings) spin- and heat-currents across the $Y$-junction, as determined respectively by the contribution to the boundary 
    interaction term in Eq.(\ref{cmh.15}) proportional to $\bar{J}_1+\bar{J}_2$ ({\bf a)}), and 
    $\propto \bar{J}_1-\bar{J}_2$ ({\bf b)}). To ease reading 
    the figure, we use full black (red) dots to denote particle (hole) like excitations within each lead. We also evidenced
    the spin polarization of the excitation entering/exiting the junction in each process. }
    \label{processes}
\end{figure}

At a glance to Eqs.(\ref{fiet.10}) and (\ref{fiet.12}) we note that, 
as implied by the $\mathbb{Z}_2$ spin symmetry of the low-energy description, no off-diagonal terms appear in the conductance
tensors. Also, we point out the different sign in the spin- and in 
the heat-current patterns determined by the boundary terms $\propto \bar{J}_1+\bar{J}_2$ and $\propto \bar{J}_1-\bar{J}_2$, 
respectively. This is determined by the different nature of the physical processes induced  
by the corresponding boundary interaction terms in Eq.(\ref{cmh.15}).  Indeed, while  the boundary interaction 
term $\propto \bar{J}_1+\bar{J}_2$ commutes with the total magnetization along the $z$ direction, the term $\propto \bar{J}_1-\bar{J}_2$ violates the associated conservation law. 
Specifically, as we depict in 
Fig.\ref{processes} {\bf a)}, the boundary interaction term $\propto \bar{J}_1+\bar{J}_2$ gives rise to spin-preserving transmission or reflection of excitations. In our conventions, the magnetization current is positive when a spin up exits  or a spin down enters the junction, negative when a spin up enters or a spin down exits the junction.
At a finite Zeeman field $\delta H_\lambda$ applied to lead $\lambda$, a  nonzero transmission amplitude from lead-$\lambda$ 
to leads $\lambda \pm 1$ provides a finite, positive current in the corresponding leads.
At the same time, the finite transmission amplitude comes along with a reduction in the reflection coefficient back into lead-$\lambda$ (equal to unity in the disconnected junction limit), thus providing a nonzero value for the current flowing through lead-$\lambda$. Since the current flows toward the junction, the corresponding conductance is negative in this case. 
At variance, in Fig.\ref{processes} {\bf b)} we show the elementary physical processes determined by the boundary interaction term $\propto \bar{J}_1-\bar{J}_2$. In this case the transmission of spin waves from lead-$\lambda$ to leads $\lambda \pm 1$ is anomalous, in the sense that it violates spin-$z$
conservation. For instance, an incoming spin up (black dot, in 
the figure) is transmitted as a spin down (red dot) toward either lead. This implies a sign change in the corresponding spin current (and, of course, in the conductance) that is opposite to the previous case, in agreement with 
the result of Eqs.(\ref{fiet.10},\ref{fiet.12}).
Importantly, within lead-$\lambda$ itself the anomalous reflection results 
in a backscattering with opposite spin, which would be 
analogous to Andreev reflection in the fermionic description. As in the previous cases, such processes determine a negative value of the current flowing through lead-$\lambda$. 
The previous argument elucidates the relative signs in 
the various contributions to the current pattern determined by the two boundary interaction terms. Instead, the heat current is only sensitive to whether a magnon is reflected back within the same lead it comes from, or transmitted toward a different lead, 
regardless of its spin. Therefore, it makes no difference whether the process involves spin-conserving or spin-flip transmission/reflection.

\noindent
Incidentally, we note that, in the symmetric case $J_1=J_2$, all the (Jordan-Wigner fermion) charge (that is, 
physical spin) conservation 
violating interaction terms disappear from $H_{K,f}$ in Eq.(\ref{cmh.15}). Accordingly, the spin conductance tensor
takes the only possible form allowed by the spin conservation and by the symmetry between the three channel. 
Specifically, we obtain 

\begin{eqnarray}
I_{S,\lambda} (x) &=&2 \bar{J}^2 {\cal A}_S (T)   \{ 2 \delta H_\lambda - \delta H_{\lambda +1} - 
\delta H_{\lambda -1} \} 
\label{fiet.12yy}  \\
I_{Q,\lambda} (x) &=& 4 \bar{J}^2  \: {\cal A}_Q (T) 
\{ - 2 \delta T_\lambda + \delta T_{\lambda + 1} + \delta T_{\lambda -1} \} 
\;\; , \nonumber 
\end{eqnarray}
\noindent
with $\bar{J} = \bar{J}_1 = \bar{J}_2$. In this case, as already noted in Refs.\cite{Buccheri2022,Giuliano2022}, 
the Wiedemann-Franz law is recovered, as a direct consequence of the absence of the anomalous scattering processes at the junction. 

Consistently to our previous RG analysis, we improve the results
of  Eqs.(\ref{fiet.10},\ref{fiet.12}) by substituting the bare couplings 
with the running ones, $\bar{J}_1 (D),\bar{J}_2 (D)$, which dynamically depend on the running energy scale $D$ (which we 
eventually identify with $k_B T$). To determine their functional dependence on $D$ at given values of the system 
parameters, one should integrate Eqs.(\ref{cmh.9}). However, to qualitatively illustrate the main 
trend of the conductance tensors when entering the nonperturbative regime, we may employ
the simplified Eqs.(\ref{ap.b.35}), thus getting 

\beq 
\bar{J}_{1,2} (T) \approx  \frac{\bar{J}_{1,2} (D_0)}{1 + \frac{ 3 \bar{J}_{1,2} (D_0) }{\pi v} \ln \left(\frac{k_B T}{D_0} \right)}
\:\:.
\label{fiet.xx1}
\eneq
\noindent
Eqs.(\ref{fiet.xx1}) imply a logarithmic increase of both running couplings, until $T$ reaches the Kondo temperature in 
whichever channel has the largest bare boundary coupling ($\bar{J}_{1,2} (D_0)$). 
This implies that, on lowering $T$ toward the Kondo temperature, $I_{S,\lambda} (x) \to 4 \bar{J}^2 {\cal A}_S (T) 
\delta H_\lambda$. This is, in fact, consistent with the results that we derive in the next Section for the 
${\cal I}_1$ and ${\cal I}_2$ FPs.

\section{Spin and heat current at the Ising-like fixed point}
\label{s_fixed} 

In Appendix \ref{islike} we derive in detail the field-theory description of the   ${\cal I}_1$ and of 
the ${\cal I}_2$ FPs of the phase diagram of the junction. In particular, we introduce the
formal description of the system at the FP in terms of a pair of triples of  chiral
real-fermionic field operators, $\{ \bar{\xi}_\lambda (x) , \bar{\zeta}_\lambda (x)\}$, so that 
the FP Hamiltonian and the current density operators take the expression
listed in Eqs.(\ref{srois.16x}). As a result, we note that, while the FP Hamiltonian apparently takes a form similar to 
 what we had at the DFP, the current operators are substantially different from what they 
 are at the DFP, which we report in  Eq.(\ref{fiet.4}). Such a difference has important 
 consequences for the transport properties of the system at the ${\cal I}_1$ FP, 
 as we are going to discuss next.

Within Keldysh approach, the currents are given by

 \begin{eqnarray}
 I_{S,\lambda} (x) &=& - \hbar v \lim_{x' \to x} \:
 \sum_{a= \pm 1} \: 
  \int \: \frac{d \omega}{2 \pi} \times   \label{ish.3}  \\
  && \{G^\lambda_{(+,-)} (ax,ay ; \omega) 
   +  G^\lambda_{(-,+)} (ay,ax;\omega) \} 
  \nonumber \\
  I_{Q,\lambda}(x) &=& - i \hbar v^2  \lim_{x' \to x} \: \: \sum_{a= \pm 1} \: \partial_y 
  \: \int \: \frac{d \omega}{2 \pi} \times \nonumber \\
&& \{ G^\lambda_{(+,-)} (ax,ay ; \omega ) + G^\lambda_{(-,+)} (ay,ax;\omega) \} 
  \:\: ,
\nonumber
  \end{eqnarray}
  \noindent
 with the $G^\lambda_{(\eta,\eta')} (x,x';\omega)$ being the Keldysh-Green functions of the 
 $\bar{\psi}_\lambda$-field ($=\frac{\bar{\xi}_\lambda + i \bar{\zeta}_\lambda}{\sqrt{2}}$)
  in frequency space.  An important difference with respect to the DFP is
 that, at the ${\cal I}_1$ FP, we already obtain a nonzero result for the spin current, even without
 considering (irrelevant)  boundary perturbations of the FP Hamiltonian. 
 Indeed, using 
 Eqs.(\ref{fiet.7}) for the Keldysh-Green functions, we obtain 
 
 \beq
 I_{S,\lambda} (x) = -\frac{1}{2 \pi } \intop_{-D_0}^{D_0} \: d \omega \: \{ f(-\omega + \delta H_\lambda ) - f (\omega - \delta H_\lambda ) \}
 \:\: . 
 \label{ish.4}
 \eneq
 \noindent 
In Eq.(\ref{ish.4}) we have explicitly denoted the high-frequency cutoff $D_0$, which is required to reguralize our following calculations). 
 To recover the correct result, 
 we have, indeed, to perform all the calculations by keeping $D_0$ finite and by sending $D_0 \to \infty$ only
 at the end of the whole derivation. As a result, we obtain 
 
 \beq
 I_{S,\lambda} (x) = - \frac{\hbar \delta H_\lambda}{\pi} 
 \:\: . 
 \label{ish.5}
 \eneq
 \noindent
 At variance, for the heat current we obtain $I_{Q,\lambda} (x)=0$.  Remarkably, while for both
 the spin, and the heat, transport there is no conductance between different chains,
  from Eq.(\ref{ish.5}) we note a saturation of the intra-channel
 spin conductance, up to  twice the maximum value it should take in a ballistic spin conductor. 
 In fact, such a behavior  corresponds to what happens at a junction between a p-wave superconductor
 in its topological phase and a spinless quantum wire \cite{Fidkowski2012,Affleck2013,Affleck2014,Giuliano2022}. 
 In that case, the saturation of the electric conductance is a result of the resonant (at the Fermi level) 
 Andreev scattering amplitude due to the  full hybridization between the degrees of 
 freedom of the metallic lead and the zero-energy Majorana mode emerging at the NS interface. In our case,
 saturation of the spin conductance is due to the hybridization between the degrees of freedom of lead $\lambda$ and the 
 corresponding Klein factor. In this respect, we also point out how, since for each incoming particle/hole from the lead a single hole/particle is Andreev backscattered within the same lead, no net energy flow is realized across the  lead itself, which motivates the result of a zero thermal current within each lead. 
 
 To move slightly away the ${\cal I}_1$ (or the ${\cal I}_2$). FP, we add to the FP Hamiltonian 
 the leading boundary interaction term allowed by symmetry at that FP, $H_{{\cal I}_1}$,
 given by
  
 \beq
 H_{{\cal I}_1} =  i \{ {\cal A} \prod_{\lambda=1}^3 \bar{\xi}_\lambda(0) + 
 {\cal B} \sum_{\lambda =1}^3 \bar{\xi}_\lambda (0) \bar{\zeta}_{\lambda +1} (0)  
 \bar{\zeta}_{\lambda +2} (0) \} 
 \;\; , 
 \label{ishtar.1x}
 \eneq
 \noindent
 with ${\cal A}$ and ${\cal B}$ interaction strengths whose expressions in terms of 
 the lattice Hamiltonian parameters we provide in Eqs.(\ref{srois.14}). $H_{{\cal I}_1}$ has 
 scaling dimension $\frac{3}{2}$ and is, therefore, an irrelevant perturbation. 
 Being $H_{{\cal I}_1}$ irrelevant, we expect it to provide corrections 
 to the FP currents that go to zero as $T\to 0$. To compute them, we 
 proceed as we have done at the DFP, that is, we employ Eqs.(\ref{ish.3}) to compute the    
 $G_{(\eta,\eta')}^\lambda(x,x';t)$  to second-order
in $H_{{\cal I}_1}$. As a result, we obtain 

\begin{eqnarray}
&& G^\lambda_{(\eta,\eta')} (x,x';t) =  g^\lambda_{(\eta,\eta')} (x,x';t) + \delta G^\lambda_{(\eta,\eta')} (x,x';t) \nonumber \\
&& \delta G^\lambda_{(\eta,\eta')} (x,x';t)  \approx  - \frac{1}{2} \: \sum_{\eta_1,\eta_2} \: \int \: d t_1 \: d t_2 \: 
\eta_1 \eta_2 \: \times 
\label{ish.6}\\
&&  \langle {\bf T}_K \bar{\psi}_\lambda (x,t,\eta) \bar{\psi}_\lambda^\dagger (x',0,\eta') H_{{\cal I}_1} (t_1,\eta_1) 
H_{{\cal I}_1} (t_2,\eta_2) \rangle 
\:\: . \nonumber 
\end{eqnarray}
\noindent
Collecting together the result of Eq.(\ref{ish.5}) and what we get from Eq.(\ref{ish.6}), we eventually
obtain 

\begin{eqnarray}
I_{S,\lambda} (x) &=&  - \frac{\delta H_\lambda}{\pi} 
+  \kappa_S T \{ \Sigma_S  \delta H_\lambda + \Lambda_S [ \delta H_{\lambda +1} + \delta H_{\lambda-1} ] \} 
\nonumber \\
I_{Q,\lambda} (x) &=& \kappa_Q T^2 \{\Sigma_Q \delta T_\lambda + \Lambda_Q [\delta H_{\lambda +1} + \delta H_{\lambda -1} ]\}
\;\; , 
\label{ishtar.1}
\end{eqnarray}
\noindent
with $\kappa_S$ and $\kappa_Q$ being over-all constants (in the $T \to 0$ limit) and 
$\Sigma_S,\Lambda_S,\Sigma_Q,\Lambda_Q$ being numerical coefficients. 
 
The above results readily allow us to identify the ${\cal I}_1$ (as well as the ${\cal I}_2$, of course) FP with 
the $N=3$  $D^N$ FP of a junction of interacting quantum wires \cite{Giuliano2022}. The  perfect intra-channel spin
 conductance is due to the FP interplay between the physical processes shown in Fig.\ref{processes}. Indeed, while, at the level of algebraic structure, the theory underneath the ${\cal I}_1$ FP shows no differences with respect to the one describing the DFP
 (in both case the three leads are disconnected), in fact, moving from the DFP to the ${\cal I}_1$ FP corresponds to 
 a change in the boundary conditions of the JW fermions from open to Andreev.
  It is, therefore, not surprising that all the conductances are zero but the intra-channel spin-conductance, whose 
  value is actually twice as large as one would get for a perfectly, spin-conducting wire. 
  At the same time, the Lorenz ratio at this FP is vanishing, $\mathcal{L}=0$,
   indicating the strongest violation of the magnetic WF. 

Subleading processes encoded in the ``residual'' boundary interaction of 
Eq.(\ref{strois.17}) modify the FP result.  Due to the irrelevance (in the 
RG sense) of all the operators contributing the right-hand side of Eq.(\ref{strois.17}) we,
however,  find that all the corrections due to the residual boundary operators would be negligible
at low enough energy scales, leaving only the FP result discussed above. 

As we discuss in the following, the 
symmetric coupling case $J_1 = J_2$ determines a special flow direction in 
parameter space, along which the system flows towards a TK-like FP.

\section{Spin and heat current at the Topological Kondo fixed point}
\label{sicatopo}

 In the symmetric coupling case $J_1=J_2$ our system flows
 toward the $N=3$  TK FP. In Appendix \ref{ren_sym}, we resort to bosonization approach to reformulate 
 the corresponding FP theoretical description of the system. Here we just summarize the main steps but discuss
 in detail the results.  Starting   from the  bulk Hamiltonian expressed in terms of the unfolded fields, that is,   
 
 \begin{eqnarray} \label{st.1}
 H_0 &=& -i   \hbar v \: \sum_{\lambda = 1}^3 \: \int_0^\ell \: d x \: \times 
 \\ && \{\psi_{R,\lambda}^\dagger (x) \partial_x \psi_{R,\lambda} (x)
 -  \psi_{L,\lambda}^\dagger (x) \partial_x \psi_{L,\lambda}(x) \} 
 \:\: , \nonumber
 \end{eqnarray}
 \noindent 
we bosonize the chiral fermionic fields by introducing two triples of chiral bosonic fields, $\{\varphi_{R,\lambda} (x) , \varphi_{L,\lambda} (x) \}$,
so to set 

\begin{eqnarray}
 \Gamma_\lambda\: \psi_{R,\lambda}(x) &=&  e^{i \sqrt{4\pi} \varphi_{R,\lambda}(x) } \nonumber \\
 \Gamma_\lambda\: \psi_{L,\lambda} (x) &=& e^{i \sqrt{4\pi} \varphi_{L,\lambda} (x) } 
\:\:, 
\label{st.2}
\end{eqnarray}
\noindent
with the $\Gamma_\lambda$ being additional Klein factor, required to preserve the correct (anti)commutation relations 
between fields living on different arms of the junction \cite{Tsvelik2013,Crampe2013}. The canonically conjugate nonchiral fields 
$\phi_\lambda (x),\theta_\lambda (x)$
are therefore defined as

\begin{eqnarray}
\phi_\lambda (x ) &=& \varphi_{R,\lambda} (x) + \varphi_{L,\lambda} (x) \nonumber \\
\theta_\lambda (x) &=& \varphi_{R,\lambda}(x) - \varphi_{L,\lambda}(x) 
\:\: . 
\label{st.3}
\end{eqnarray}
\noindent
In terms of the canonical fields in Eqs.(\ref{st.3}) we respectively obtain, for the spin- and for the heat-currents, the
expressions 

\begin{eqnarray}
{\cal J}_{S,\lambda}^z (x) &=& \frac{\hbar v}{\pi}  \partial_x \phi_\lambda (x) \nonumber \\
{\cal J}_{Q,\lambda}(x) &=& \frac{\hbar v^2}{\pi^2}  \partial_x \phi_\lambda (x) \partial_x \theta_\lambda (x) 
\:\: . 
\label{st.4}
\end{eqnarray}
\noindent
Within bosonization framework, we note that, at the DFP,  
 we get the boundary conditions $\sqrt{4 \pi} \varphi_{R,\lambda} (0) = 
\sqrt{4\pi} \varphi_{L,\lambda} (0) + 2 \pi n_\lambda$, with integer $n_\lambda$. This implies that $\theta_\lambda (0)$ 
is a constant. As a result, we recover the boundary interaction in bosonic coordinates, $H_\Delta$ as 
\beq
H_B = - 4 \bar{J} \: \sum_{\lambda=1}^3 \: \cos [\sqrt{\pi} (\phi_\lambda (0) - \phi_{\lambda +1} (0))]  \:\: .
\label{st.6}
\eneq
\noindent
To access the TK FP it is more effective to switch to the bosonic center-of-mass and 
relative coordinate, $\Phi (x),\chi_1 (x) , \chi_2(x)$, together with the corresponding
dual fields $\Theta (x),\omega_1 (x),\omega_2 (x)$,  defined by 

\beq
\left[\begin{array}{c} \Phi (x) \\ \chi_1 (x ) \\ \chi_2 ( x ) \end{array} \right] 
= \left[ \begin{array}{ccc} \frac{1}{\sqrt{3}} & \frac{1}{\sqrt{3}} & \frac{1}{\sqrt{3}} \\ \frac{1}{\sqrt{2}} & - \frac{1}{\sqrt{2}} & 0 \\
\frac{1}{\sqrt{6}} & \frac{1}{\sqrt{6}} & - \frac{2}{\sqrt{6}} \end{array} \right] 
\left[ \begin{array}{c} \phi_1 ( x ) \\ \phi_2 ( x ) \\ \phi_3 ( x ) \end{array} \right] \;\;,
\label{st.7p}
\eneq
\noindent
and by

\beq
\left[\begin{array}{c} \Theta (x) \\ \omega_1 (x ) \\ \omega_2 ( x ) \end{array} \right] 
= \left[ \begin{array}{ccc} \frac{1}{\sqrt{3}} & \frac{1}{\sqrt{3}} & \frac{1}{\sqrt{3}} \\ \frac{1}{\sqrt{2}} & - \frac{1}{\sqrt{2}} & 0 \\
\frac{1}{\sqrt{6}} & \frac{1}{\sqrt{6}} & - \frac{2}{\sqrt{6}} \end{array} \right] 
\left[ \begin{array}{c} \theta_1 ( x ) \\ \theta_2 ( x ) \\ \theta_3 ( x ) \end{array} \right] \:\: .
\label{st.7}
\eneq
\noindent
At the DFP, the open boundary conditions imply $\partial_x \phi_\lambda (x=0) = 0$, 
$\forall \lambda=1,2,3$. Inverting Eqs.(\ref{st.7}), we  see that the right-hand side of 
Eq.(\ref{st.6}) only depends on 
 $\chi_1 (0)$ and $\chi_2 (0)$. As we show in Appendix \ref{ren_sym}, this implies that the strongly coupled limit is 
  recovered by pinning both $\chi_1(x)$ and $\chi_2(x)$ at $x=0$, that is, by imposing Dirichlet boundary conditions
 on both fields, which correspond to Neumann boundary conditions over the correspondingl dual fields. We therefore
 obtain that the strongly coupled FP is described by the boundary conditions
 
 \beq
 \partial_x \Phi (x=0) = \partial_x \omega_1 (x=0 ) = \partial_x \omega_2 (x=0 ) = 0 
 \:\: . 
 \label{st.8}
 \eneq
 \noindent
Given Eq.(\ref{st.8}), it is a simple exercise to compute the 
spin- and the heat -current pattern at the TK FP by resorting, for instance, to 
the splitting matrix method \cite{Eriksson2014,Giuliano2015,Buccheri2022,Giuliano2022}.
As a result, one obtains that, to linear order in the transverse magnetic field and in the temperature
bias, one gets 

\begin{eqnarray}
{\cal I}_{S,\lambda}(x) &=& \sum_{\lambda_1=1}^3 \: {\bf G}^{S,{\rm TK}}_{\lambda,\lambda_1} \: \delta H_{\lambda_1} \nonumber \\
{\cal I}_{Q,\lambda} (x) &=&  \sum_{\lambda_1=1}^3 \: {\bf K}^{Q,{\rm TK}}_{\lambda,\lambda_1} \: \delta T_{\lambda_1} 
\:\: , 
\label{st.9}
\end{eqnarray}
\noindent
with 
\beq
{\bf G}^{S,{\rm TK}}  = -\frac{\hbar}{3 \pi} \: \left[\begin{array}{ccc} 2 &-1 &-1 \\ -1 & 2 & -1 \\ -1 & -1 & 2 \end{array} \right]
\;\; , 
\label{st.10}
\eneq
\noindent
and 

\beq
{\bf K}^{S,{\rm TK}} =  -\frac{2 \pi k_B^2 T}{27\hbar} \: \left[\begin{array}{ccc} 2 &-1 &-1 \\ -1 & 2 & -1 \\ -1 & -1 & 2 \end{array} \right]
\;\; . 
\label{st.11}
\eneq
\noindent
Taking the ration between any matrix element of the heat- and of the spin-conductance tensor, we find a 
remarkable renormalization of the Lorenz ratio, compared to what predicted by the WF law, Eq.(\ref{LorenzRatio}). Indeed, 
at the FP we obtain  

\beq
\mathcal{L}=\frac{2 \pi^2}{9}\left(\frac{k_B}{\hbar} \right)^2 \,\; . 
\label{st.12}
\eneq
\noindent 
so that, comparing to \eqref{WF}, a factor $2/3$ appears. As 
at the ${\cal I}_1$ FP, we see that the WF law is violated at the 
TK FP, as well \cite{Buccheri2022,Giuliano2022}. However, differently from 
the ``trivial'' violation we found at the $D^3$ (${\cal I}_1$) FP, here
the combined effect of the  boundary conditions in Eq.(\ref{st.8}) and
of the charge conservation implies that the low-temperature
physics corresponds to the splitting of  an injected 
JW fermion (that is, a spin-1/2 ``spinon'' excitation)  
 into fractional spin excitations  propagating
into the other two wires (carrying 2/3 times of the incoming spin), 
and backscattered into the same wire (carrying 1/3 of the incoming spin).
This evidences a remarkable onset of the spin-heat separation, 
that is, the spin analog of the charge-heat separation discussed in 
\cite{Buccheri2022,Giuliano2022}.  By analogy with the 
``standard''  TKM, we interpret the universal renormalization
of the Lorenz ratio 
as a hallmark of the hybridization between the real fermionic
Klein factors emerging in the formal description of the junction.
In this respect, it is the spin-conserving version of the saturation 
of the spin conductance that we discussed when analyzing the ${\cal I}_1$ 
FP. Remarkably, the strict analogy between the behavior junctions 
 of three interacting quantum wires and of quantum XX spin chains
suggests the natural possibility of using the latter ones, that can 
be in principle realized in experimental devices with a quite good
level of control on the system parameters, to analogically simulate 
the theoretically rich physics of quantum wires connected to real
Majorana fermions, so far not yet unambiguously detected in 
experiments on real systems. 
 
 To conclude the discussion of the TK FP we note that, 
 consistently with the phase diagram of Fig.\ref{pha_dia}, we expect that
 the corresponding FP is unstable against inosotropies in the spin exchange 
 interactions at the junction along the $x$ and the $y$ direction in spin space.
 This means that, to keep consistent with our phase diagram, any 
 nonzero $J_1-J_2$ should induce a relevant operator taking 
 the junction from the TK to either the ${\cal I}_1$, or the ${\cal I}_2$, FP. 
 Indeed, in Appendix \ref{ren_sym} we construct the corresponding operator $\delta \tilde{H}_\Delta$  by means of a systematic implementation of the DEBC method, eventually finding 

 \beq
 \delta \tilde{H}_\Delta = -4 (\bar{J}_1-\bar{J}_2) 
 \cos \left[ \frac{2 \sqrt{\pi} \Phi (0) }{\sqrt{3}} \right]
 \:\: . 
 \label{sf.13}
 \eneq
 \noindent
 $\delta \tilde{H}_\Delta$ is a relevant operator, with scaling dimension $d_\delta J = \frac{2}{3}$: it provides 
 the relevant perturbation that takes the system out of the TK FP and triggers the flow towards 
 either one of the Ising-like FPs.

\section{Discussion and concluding remarks}
\label{concl}

 We have systematically computed the thermal and the spin-current pattern across a Y junction of XX spin chain
at varying values of the magnetic exchange interactions between the spins at the junction point. Doing so, we have been 
able to map out    the boundary phase diagram of the  junction of XX spin chain and to fully characterize the various FPs, in 
the presence of an anisotropic spin exchange 
interaction at the junction.  The junction with purely isotropic XX interactions is known to have a nontrivial FP 
analogous to a four-channel Kondo model \cite{Crampe2013}, robust again small perturbations of the coupling constants. In contrast, we have shown that
 this FP is unstable against an XY anisotropy in the spin exchange couplings. In particular, we have evidenced how, 
in the presence of the anisotropy, the system flows toward any of two equivalent Ising FPs \cite{Coleman1995,Giuliano2016,Zazunov2017}. As a consequence, the low-temperature magnetic conductance from one leg to 
another turns from the universal $2\hbar/3\pi$ to zero, while the Lorenz ratio turns from the universal $2\pi^2k_B^2/9\hbar^2$ to zero. 
Avoiding this dangerous instability requires careful control of the experiment parameters and may hinder the observability of the 
four-channel, TK physics in junctions of quantum spin chains.

 At the TK fixed point, our combined use of the charge- and of the spin-conductance for analyzing the phase diagram of the junction 
has evidenced the remarkable emergence of the spin analog of the charge-heat separation in analogous junctions made with conducting 
quantum wires \cite{Buccheri2022,Giuliano2022}. At the same time, the unique possibility of realizing the same physics of a junction of three quantum Ising chains, 
but now with spin current conservations within the (XX) leads, determines the onset of the same physics that is  expected (but still not unambigously proven) 
at a NS-junction hosting an isolated  Majorana mode. Given the possibility  hat pertinently shaped Y-junction traps are nowadays well within the reach of 
current technologies \cite{Buccheri2016}, our results apparently suggest the alternative possibility of employing simple networks, like junctions of three XX chains, 
as a controlled background to host emerging, localized Majorana modes.  

Regarding the observables, we have focused on the heat transport by spin waves only. 
In nanostructures, even electric insulators can carry gapless excitations, namely, lattice 
deformations in the form of acoustic phonon modes and spin waves. Each of these carries energy and 
contributes to the thermal-, but not to the electric-, conductance.   
In addition, we have considered purely ballistic transport. Nevertheless, the presence of impurities turns out to be an important factor for the temperature dependence of the thermal conductivity in actual spin chain materials \cite{Rozhkov2005}. 
This qualitative picture can be further refined by the inclusion of the spin-phonon coupling \cite{Louis2006} 
and of the details of the materials, a necessary step for comparing with experiments with crystalline samples \cite{Chernyshev2016,Sologubenko2000,Sologubenko2000l,Sologubenko2001,Pan2022}. 
We expect our study to be directly relevant in the context of cold-atomic circuits and simulators, prominently, of spin chains \cite{Toskovic2016,Jepsen2020}. Ising spin chains \cite{Schauss2018,Browaeys2020}, as well as XX \cite{Geier2021} and XXZ chains \cite{Franz2024} have indeed been realized using Rydberg atoms and their dipole interaction. Such synthetic systems allow, in principle,
for measuring  the spinon contribution to the thermal conductance \cite{Hirobe2017}. 

  In view of the possible practical realization of devices such as our Y junction it is also important to discuss the potentially dangerous effects of differences in the bulk and/or the boundary magnetic interaction strengths of the chains. Within our approach, different interaction strengths at the junction would result in a dependence of $J_1$ and of $J_2$ on the index $\lambda$. That would give rise to an effective anisotropy in the Kondo couplings which, in the specific realization of the
Kondo effect at a junction of spin chains, is known to be washed out along the RG trajectoris (that is, anisotropies are irrelevant for  the low-energy/temperature behavior of the system)
\cite{Tsvelik2013,Altland2013}. At variance, 
having different bulk magnetic interaction strength in the bulk of the chains would lead to a $\lambda$-depending scale 
$J_\lambda$. Within our approach, this would simply imply a dependence on $\lambda$ of the Fermi velocity ($v \to v_\lambda$) that, on its own, would merely result in an 
effective anisotropy in the boundary couplings. As we discussed above, that anisotropy has no effects for the physical properties of the Kondo FPs. Eventually, the above argument
shows how it is, in fact, not necessary to fine-tune the system parameters so to be exactly the same for all three the chains, in order to recover the physics we discuss in our paper. 
 On the other hand, the realization of an anisotropy in the exchange interaction in optical lattice, $J_1\ne J_2$ in \eqref{mh.2}, can be achieved via the application of microwave drive, within present-day capabilities. Even accounting for diffraction-limited focusing, we expect that application in the center of the junction can be achieved, because typical distances between Rydberg atoms or ions are of several to several tens of $\mu$m, when trapped in optical lattices. For these reasons, we are confident that the results proposed in this paper will be put to test in the near future. 
 
\begin{acknowledgments}
 We thank Andrea Nava for insightful discussions. 
 \end{acknowledgments} 
 
 \vspace{0,5cm}

\hfill 
\begin{center}{\bf DATA AVAILABILITY} \end{center}

\vspace{0.5cm}

The data that support the findings of this article are openly
available at \cite{zenodo} 
 
 \appendix

\section{The Green-Keldysh functions for the chiral fermionic fields }
\label{gefu}

\begin{figure}
    \centering    \includegraphics[width=\linewidth]{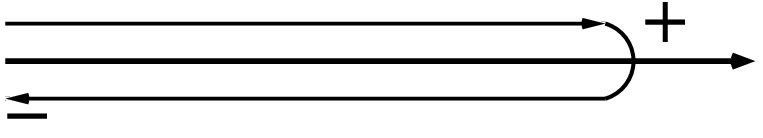}
    \caption{Sketch of the Keldysh path with the two labels $_+$ and $_-$ attached to the corresponding 
    branches. }
    \label{keld}
\end{figure}

In this Appendix we  derive the Green-Keldysh functions that we used to perform the perturbative
calculation of  the spin- and of the heat-conductance tensor in Section \ref{pefu}. 
(In this and in the following Appendixes, we set $\hbar=1$ for lightness of notation).  As a necessary preliminary step,
we derive the field-theoretical description of a single spin chain 
at a fixed temperature $T$  with an ``extra'' magnetic field $\delta H$ in the $z$ direction.
 To do so, we retain only long-wavelength excitations around the Fermi points at $\pm k_F$, with 
$-2J\cos (k_F ) + H = 0$. This implies the following expansion for the lattice fermionic fields 

\beq
c_{j,\lambda} \approx e^{ik_F j} \psi_{R,\lambda} (x_j ) + e^{-i k_F j } \psi_{L,\lambda} (x_j ) 
\;\; , 
\label{ap.c.1}
\eneq
\noindent
with $x_j = a j $ (and the lattice step $a$ set to 1 henceforth),  $\psi_{R,\lambda},\psi_{L,\lambda}$ chiral
fermionic fields described by the Hamiltonian 

\begin{eqnarray}
&& H_{\rm FT} = - i v \: \int_0^\ell \: d x \: \sum_{\lambda = 1}^3 \: \times \nonumber \\
&&  \{\psi_{R,\lambda}^\dagger (x) 
\partial_x \psi_{R,\lambda} (x) - \psi_{L,\lambda}^\dagger (x) \partial_x \psi_{L,\lambda} (x) \} 
\;\; , 
\label{ap.c.2}
\end{eqnarray}
\noindent
and the Fermi velocity $v = J \sqrt{1 - \left(\frac{H}{2J}\right)^2 }$. The open boundary conditions for 
the lattice fields at $j=0,\ell + 1$ imply

\begin{eqnarray}
&& \psi_{R,\lambda} (0 ) + \psi_{L,\lambda} (0 ) = 0 \label{ap.c.3} \\
&& e^{ik_F (\ell + 1 ) } \psi_{R,\lambda} (\ell + 1 ) + e^{-ik_F (\ell + 1) } \psi_{L,\lambda} (\ell + 1 ) = 0 
\:\; .
\nonumber 
\end{eqnarray}
\noindent
Accordingly, for $-\ell \leq x \leq \ell$, we define  a set of chiral, right-handed fermionic fields $\psi_\lambda (x)$, 
 by setting 

\beq
\psi_\lambda (x) = \Biggl\{ \begin{array}{l} \psi_R (x) \;\; , \; (0 \leq x \leq \ell) \\
- \psi_L (-x) \;\; , \; (-\ell \leq x < 0 ) \end{array}
\;\; , 
\label{ap.c.4}
\eneq
\noindent
which implies 

\beq
e^{i\delta} \psi_\lambda (\ell ) - e^{-i\delta} \psi_\lambda (-\ell ) = 0 
\:\: , 
\label{ap.c.5}
\eneq
\noindent
with $\delta = k_F (\ell + 1)$. 
As a result, the Hamiltonian in Eq.(\ref{ap.c.2}) takes the form 

\beq
H_{\rm FT} = - i v \: \intop_{-\ell}^\ell \; d x \; \sum_{\lambda = 1}^3 \: \psi_\lambda^\dagger (x) \partial_x \psi_\lambda (x) 
\:\: . 
\label{ap.c.6}
\eneq
\noindent
Adding an extra magnetic field $\delta H_\lambda$ to lead $\lambda$ results in an additional 
Hamiltonian contribution $H_{\delta H}$, given by

\beq
H_{\delta H} = - \sum_{\lambda = 1}^3 \: \delta H_\lambda  \: \intop_{-\ell}^\ell \: d x \:
\psi_\lambda^\dagger (x) \psi_\lambda (x) 
\:\: . 
\label{ap.x.1}
\eneq
\noindent
In terms of the unfolded fields,  the spin- and heat-current operators now become

 \begin{eqnarray}
  {\cal J}_{S,\lambda}^z (x,t) &=&v  \: \sum_{a = \pm 1} \: a \psi_\lambda^\dagger (ax,t) \psi_\lambda (ax,t) \nonumber \\
  {\cal J}_{Q,\lambda} (x,t) &=& -\frac{ i v^2}{2}  \: \sum_{a = \pm 1} \: \{   \psi_\lambda^\dagger (ax,t) \partial_x \psi_\lambda (ax,t) 
  \nonumber \\
  &-& (\partial_x \psi_\lambda^\dagger (ax,t) ) \psi_\lambda (ax,t) \} 
  \:\: . 
  \label{ap.c.8}
  \end{eqnarray}
  \noindent
Introducing,  over a Keldysh path as the one displayed in Fig.\ref{keld}, the labels $_+$ and $_-$ to
respectively refer to the upper and to the lower branch of the path itself, we   define 
 the single-fermion  Green's functions for lead-$\lambda$, $g^\lambda_{(\eta,\eta')} (x,x';t)$ as 

\beq
g^\lambda_{(\eta,\eta')} (x,x'; t) =  \langle {\bf T}_K \psi_\lambda (x,t,\eta) \psi_\lambda^\dagger (x',0,\eta') \rangle
\;\; . 
\label{gefu.1}
\eneq
\noindent
Thus, we obtain

\begin{eqnarray}
&& g_{(+,+)}^\lambda (x,x';t)  =  \frac{1}{2\ell} \: \sum_k \: e^{ik(x-x') - i \epsilon_k t} \: \times \nonumber \\
&& \{\theta (t) f_\lambda  (-\epsilon_k + \delta H_\lambda  )  - \theta (-t) f_\lambda  (\epsilon_k - \delta H_\lambda  ) \}
 \nonumber \\
&& g^\lambda_{(+,-)} (x,x';t)  = \frac{1}{2\ell} \: \sum_k \: e^{ik(x-x') - i \epsilon_k t} \: f _\lambda (-\epsilon_k + \delta H_\lambda ) \nonumber \\
&& g_{(-,+)}^\lambda (x,x';t)  =  - \frac{1}{2\ell} \: \sum_k \: e^{ik(x-x') - i \epsilon_k t} \: f_\lambda  (\epsilon_k - \delta H_\lambda ) \nonumber \\
&& g_{(-,-)}^\lambda (x,x';t)  =  \frac{1}{2\ell} \: \sum_k \: e^{ik(x-x') - i \epsilon_k t} \times \nonumber \\ 
&& \{\theta (-t) f_\lambda  (-\epsilon_k + \delta H_\lambda  )  - \theta (t) f_\lambda  (\epsilon_k - \delta H_\lambda  ) \} 
\:\: ,
\label{fiet.5}
\end{eqnarray}
\noindent
with $f_\lambda$ being Fermi distribution function for chain-$\lambda$. 

\noindent
Switching to the frequency domain, according to $g_{(\eta,\eta')} (x,x';\omega) = 
\int \: d t \: e^{i\omega t} \: g_{(\eta , \eta')} (x,x';t)$, we eventually get (in the large-$\ell$ limit):

 \begin{eqnarray}
&& g_{(+,+)} (x,x';\omega ) = \frac{ e^{\frac{i\omega}{v} (x-x')}}{2v} \times \nonumber \\
&&  \left\{ \tanh  \frac{\beta_\lambda (\omega - \delta H_\lambda )}{2} +\epsilon (x-x')
 \right\}
 \nonumber \\
&& g_{(+,-)} (x,x';\omega )  =   \frac{ e^{\frac{i\omega}{v} (x-x')}}{ v} \: f_\lambda (-\omega + \delta H_\lambda ) \nonumber \\
&&  g_{(-,+)} (x,x';\omega )  =   -\frac{ e^{\frac{i\omega}{v} (x-x')}}{v} \: f_\lambda (\omega - \delta H_\lambda ) \nonumber \\ 
 && g_{(-,-)} (x,x';\omega )  =   \frac{ e^{\frac{i\omega}{v} (x-x')}}{2v} \times \nonumber \\
&& \left\{ \tanh \frac{\beta_\lambda (\omega - \delta H_\lambda )}{2} - \epsilon (x-x')
 \right\} \nonumber \\
 \:\: . 
 \label{fiet.7}
 \end{eqnarray}
 \noindent
The Green-Keldysh functions in Eqs.(\ref{fiet.7}) are what we have used in Section \ref{pefu}   to 
compute the spin- and the heat -current pattern perturbatively in the Kondo boundary interaction. 
 
\section{Mathematical derivation of the 
RG equations for the running Kondo parameters}
\label{RGK}

In this Appendix we derive the RG equations for the running boundary couplings
in the Kondo interaction Hamiltonian $H_{K,f}$. To do so, we  write the 
partition function of the model as a path integral at imaginary times and  
systematically implement the RG procedure, thus computing the corrections
to the interaction strengths arising from an infinitesimal decrease in the cutoff reference scale $D$ 
\cite{Hewson1993}. As we perform our derivation here without linearizing the dispersion
relation around the Fermi points, we are able to fully recover the dependence of the corresponding 
Kondo temperature depends on the system parameters \cite{Giuliano2016}.

Within  imaginary time formalism, the Euclidean action takes the form ${\cal S} = {\cal S}^{(0)}
+ {\cal S}_K$, with 

\begin{eqnarray}
{\cal S}^{(0)} &=& \sum_{\lambda = 1}^3 {\cal S}_\lambda^{(0)} \label{ap.b.1} \\
{\cal S}^{(0)}_\lambda &=& \int_0^\beta \: d \tau \: \Biggl\{ \sum_{j=1}^L c_{j,\lambda}^\dagger (\tau) [ \partial_\tau + H]  c_{j,\lambda} (\tau) 
 \\
&&- J \sum_{j=1}^L [c_{j,\lambda}^\dagger (\tau) c_{j+1,\lambda} (\tau) + c_{j+1,\lambda}^\dagger (\tau) c_{j,\lambda} (\tau) ] \Biggr\}
\nonumber \\
{\cal S}_K &=& \left( \frac{J_1+J_2}{2} \right) \: \sum_{\lambda=1}^3 \: \int_0^\beta \: d \tau \:
\{{\cal O}_1^\lambda (\tau) + [ {\cal O}_1^\lambda (\tau)]^\dagger \}  \\
&&+ 
 \left( \frac{J_1-J_2}{2} \right) \: \sum_{\lambda=1}^3 \: \int_0^\beta \: d \tau \:
\{{\cal O}_2^\lambda (\tau) + [ {\cal O}_2^\lambda (\tau)]^\dagger \} 
\:\: , 
\nonumber 
\end{eqnarray}
\noindent
with the operators ${\cal O}_{1,2}^\lambda (\tau)$   defined as 

\begin{eqnarray}
{\cal O}_1^\lambda (\tau) &=& [ - i \eta_\lambda (\tau) \eta_{\lambda + 1} (\tau) ] 
[ - i c_{1,\lambda}^\dagger  (\tau)  c_{1,\lambda + 1} (\tau) ] \nonumber \\
{\cal O}_2^\lambda (\tau) &=& [ - i \eta_\lambda (\tau) \eta_{\lambda + 1} (\tau) ] 
[ - i c_{1,\lambda} (\tau) c_{1,\lambda + 1} (\tau) ] 
\:\: . 
\label{ap.b.2}
\end{eqnarray}
\noindent
To derive the RG equations for $J_1$ and $J_2$, we first compute
 $\langle {\cal O}_1^\lambda (\tau)\rangle$ and $\langle {\cal O}_2^\lambda (\tau) \rangle$
up to second order in the boundary interaction strengths and, then, we reabsorb the
second order contributions that dynamically depend on $D$ in a renormalization of
the running couplings themselves. To do so, we need the explicit formulas for the single JW fermion 
and for the Klein factor correlation functions in the noninteracting theory ($J_1=J_2=0$). Over an 
$\ell$-site chain, these  are given by 

\begin{eqnarray}
&&  \langle {\bf T}_\tau c_{1,\lambda} (\tau) c_{1,\lambda'}^\dagger (\tau' ) \rangle = 
 \delta_{\lambda,\lambda'} \: g^\lambda (\tau - \tau' ) =\nonumber \\
&& \frac{2}{\ell+1} \: \sum_k \sin^2 (k)\{  [1-f_\lambda (\epsilon_k ) ] e^{- \epsilon_k (\tau - \tau' ) } \theta (\tau - \tau' ) \nonumber \\
&& - 
f_\lambda (\epsilon_k ) e^{-\epsilon_k (\tau - \tau' ) } \theta (\tau' - \tau ) \} 
\:\: , 
\label{ap.b.12}
\end{eqnarray}
\noindent
with ${\bf T}_\tau$ being the imaginary time ordering operator and
 $f_\lambda$ being the Fermi distribution function for chain-$\lambda$, and by

\beq
\langle {\bf T}_\tau \eta_\lambda (\tau) \eta_{\lambda'} (\tau') \rangle = 
\delta_{\lambda , \lambda'} \epsilon (\tau - \tau' ) 
\;\; , 
\label{ap.b.4}
\eneq
\noindent
with $\epsilon (\tau )$ being the sign function. In terms of the Green's functions in 
Eqs.(\ref{ap.b.12},\ref{ap.b.4}) and up to second order in $J_1$ and $J_2$, we obtain:

\begin{eqnarray}
  \langle {\cal O}_1^\lambda (\tau) \rangle &=&  \left(\frac{J_1+J_2}{2} \right) \: \frac{1}{\beta} \: \sum_{i\omega} \: g^\lambda (i\omega)
g^{\lambda+1} ( i\omega) \nonumber \\ &-&   \left(\frac{J_1+J_2}{2}\right)^2 \frac{1}{\beta^3} \: \sum_{i\omega_1,i\omega_2} 
\: g^\lambda(i\omega_1) g^{\lambda + 1 } (i\omega_2) \: \times \nonumber \\
&&  \sum_{i\Omega} \left[ \epsilon (i\Omega - i \omega_1 ) 
\epsilon (i \Omega - i \omega_2 ) \tilde{g}^{\lambda + 2} (i\Omega) \right] \nonumber \\
&-& \left(\frac{J_1- J_2}{2}\right)^2 \frac{1}{\beta^3} \: \sum_{i\omega_1,i\omega_2} 
\: g^\lambda(i\omega_1) g^{\lambda + 1 } (i\omega_2) \: \times \nonumber \\
&&  \sum_{i\Omega} \left[ \epsilon (i\Omega - i \omega_1 ) 
\epsilon (i \Omega - i \omega_2 ) \tilde{g}^{\lambda + 2} (-i\Omega) \right]
\;\; , \nonumber \\ 
\label{ap.b.21}
\end{eqnarray}
\noindent
and  

\begin{eqnarray}
\langle {\cal O}_2^\lambda (\tau) \rangle_2 &=&  -\left(\frac{J_1-J_2}{2} \right) \: \frac{1}{\beta} \:
 \sum_{i\omega} \: g^\lambda (i\omega)
g^{\lambda+1} ( - i\omega) \nonumber \\
&-& \frac{J_1^2-J_2^2}{4} \frac{1}{\beta^3} \: \sum_{i\omega_1,i\omega_2} 
\: g^\lambda (i\omega_1) g^{\lambda + 1} (-i\omega_2) \: \times \nonumber \\
&& \sum_{i\Omega} \left[ \epsilon (i\Omega - i \omega_1 ) 
\epsilon (-i\Omega + i \omega_2 ) \tilde{g}^{\lambda + 2} (-i\Omega) \right] \nonumber \\
 &-& \frac{J_1^2-J_2^2}{4} \frac{1}{\beta^3} \: \sum_{i\omega_1,i\omega_2} 
\: g^\lambda (i\omega_1) g^{\lambda + 1} (-i\omega_2) \: \times \nonumber \\
&& \sum_{i\Omega} \left[ \epsilon (i\Omega - i \omega_1 ) 
\epsilon (-i\Omega + i \omega_2 ) \tilde{g}^{\lambda + 2} (i\Omega) \right]
\:\: , \nonumber \\
\label{ap.b.22}
\end{eqnarray}
\noindent
with $\omega_1,\omega_2$ ($\Omega$) denoting fermionic (bosonic) Matsubara frequencies, and 

\begin{eqnarray}
g^\lambda ( i\omega ) &=& \int_0^\beta \: d \tau \: e^{ i \omega \tau} \: g^\lambda (\tau) = 
- \frac{2}{\ell+1} \; \sum_k \: \left[ \frac{\sin^2 (k)}{i\omega - \epsilon_k} \right] \nonumber \\
&=& \frac{i}{J} \: \sqrt{1-\left(\frac{i\omega - H}{2J} \right)^2}  \nonumber \\
\epsilon (i\omega) &=& \frac{1}{2} \: \intop_{-\beta}^\beta \: d \tau \: e^{i\omega \tau} \: \epsilon (\tau) = 
- \frac{2}{i\omega} \nonumber \\
\tilde{g}^\lambda (i\Omega) &=& \frac{1}{\beta} \sum_{i\omega}
g^\lambda (i\omega) \epsilon (i\Omega - i\omega) \nonumber \\
&=& 
 \frac{2}{\ell+1} \: \sum_k \: \left[ \frac{\sin^2 (k) \tanh \left(\frac{\beta \epsilon_k}{2} \right)}{
 i\Omega - \epsilon_k} \right]
 \:\: . 
\label{ap.b.13}
\end{eqnarray}
\noindent
 Now, following the recipe of Ref.\cite{Giuliano2016}, we recover the RG equations for the running couplings 
${\cal J}_g = \frac{J_1+J_2}{2}$ and ${\cal J}_u = \frac{J_1-J_2}{2}$ by
first trading the sums over $i\omega_2$ for integrals and by introducing a pertinent cutoff $D$, according to 

\beq
\frac{1}{\beta} \: \sum_{i\omega_2} \longrightarrow \: \intop_{-D}^D \: \frac{d \omega_2}{2 \pi}
\:\: . 
\label{ap.b.29}
\eneq
\noindent
Next, we rescale the cutoff $D \to D-\delta D$ and renormalize the running coupling accordingly:
${\cal J}_{g,u} \longrightarrow {\cal J}_{g,u} + \delta {\cal J}_{g,u} (D)$. Finally, we divide by $\delta D$ the
parameter renormalization. As a result, in the limit in which we assume that the relevant frequencies 
are $\ll D$, we obtain 

\begin{eqnarray}
\frac{d {\cal J}_g}{d D} &=& -\frac{6}{\pi J D} \: \varphi (D,H) \: \{{\cal J}_g^2 + {\cal J}_u^2\} \nonumber \\
\frac{d {\cal J}_u}{d D} &=&- \frac{6}{\pi J D} \: \varphi (D,H) \: 2 {\cal J}_g \: {\cal J}_u 
\:\: ,
\label{ap.b.30}
\end{eqnarray}
\noindent
with 

\begin{eqnarray}
\varphi (D,H) &=& \Re e \sqrt{1+\left(\frac{D+iH}{2J}\right)^2}  \nonumber \\
&=& \left\{ 1 + \frac{D^2-H^2}{2J^2} + \left[ \frac{D^2+H^2}{4 J^2} \right]^2 \right\}^\frac{1}{4} 
\: \times \nonumber \\
&&  \cos \left[ \frac{1}{2} {\rm atan} \left( \frac{2 HD}{4 J^2 + D^2 - H^2} \right) \right] 
\:\: . 
\label{ap.b.31}
\end{eqnarray}
\noindent
Eqs.(\ref{ap.b.30}) decouple, once written in terms of the original running couplings $J_{1,2} (D)$. 
In particular, one gets 

\beq
\frac{ d J_{1,2} }{d D} = - \frac{6}{\pi J D } \varphi (D,H) J_{1,2}^2 
\;\; , 
\label{ap.b.32}
\eneq
\noindent
which implies 

\beq
J_{1,2} (D) = \frac{J_{1,2} (D_0)}{1 + \frac{ 6 J_{1,2} (D_0)}{\pi J } \: \intop_{D}^{D_0} \: \left[ 
\frac{\varphi (u,H)}{u} \right] d u } 
\;\; .
\label{ap.b.33}
\eneq
\noindent
Eq.(\ref{ap.b.33}) implies, for the Kondo temperatures associated to the running couplings, the
expressions $T_{k,(1,2)}$, defined as 

\beq
\frac{ \pi J}{6 J_{1,2} (D_0)} = \intop_{k_B T_{k,(1,2)}}^{D_0} \: \left[ \frac{\varphi (u,H)}{u} \right] d u 
\:\: . 
\label{ap.b.34}
\eneq
\noindent
A numerical integration of Eqs.(\ref{ap.b.32}) is straightforward and would yield, as a byproduct, the 
full dependence of $T_{K,(1,2)}$ on $H$. However, we may find a reasonable analytic approximation 
by considering that the relevant contribution to the diverging RG flow of the running 
couplings comes from the $D \to 0$ limit. Accordingly, for the purpose of estimating the 
Kondo temperature we may approximate Eqs.(\ref{ap.b.32}) as 

\beq
\frac{d J_{1,2}}{d D} \approx - \left[ \frac{6 \varphi (0,H)}{\pi J D} \right] \: J_{1,2}^2 
\;\; , 
\label{ap.b.35}
\eneq
\noindent
which yields

\beq
T_{K,(1,2)} \approx D_0 e^{-\frac{\pi J}{6 J_{1,2} (D_0 ) \left[1 - \frac{H^2}{4J^2} \right]^\frac{1}{2} } }
\;\; .
\label{ap.b.36}
\eneq
\noindent
Eq.(\ref{ap.b.36}) implies a finite-$H$ decrease of $T_{K,(1,2)}$  from the symmetric limit $H=0$
depending on $1-\frac{H^2}{4J^2}$, which is consistent with the results derived in 
the XX limit in the lattice model studied in Ref.\cite{Giuliano2016}. 
Starting from the RG Eqs.(\ref{ap.b.32}) we recover the phase diagram discussed in 
the main text.

\section{Effective theory for the Ising-like fixed points}
\label{islike}

When $J_1 \neq J_2$ the RG flow takes our system toward a strongly coupled FP to be identified with 
the two-channel Kondo fixed point  of a Y junction of three critical Ising chain \cite{Tsvelik2013,Giuliano2013,Giuliano2020}. To 
formally describe it,   we introduce real chiral  fermionic fields 
$\xi_{R(L),\lambda} (x) , \zeta_{R(L),\lambda} (x)$, that are related to $\psi_{R(L),\lambda} (x)$ by means
of the relations 

 \begin{eqnarray}
 \psi_{R(L),\lambda} (x,t) &=& \frac{1}{\sqrt{2}} \{\xi_{R(L),\lambda} (x,t) + i \zeta_{R(L),\lambda} (x,t) \}   \label{ap.c.9} \\
 \psi_{R(L),\lambda}^\dagger (x,t) &=& \frac{1}{\sqrt{2}} \{\xi_{R(L),\lambda} (x,t) - i \zeta_{R(L),\lambda} (x,t) \} 
 \:\: .
\nonumber 
 \end{eqnarray}
 \noindent
 In terms of  $\xi_{R(L),\lambda}^\dagger$ and of $\zeta_{R(L),\lambda}^\dagger (x)$. we obtain 

\begin{eqnarray}
&& H_0  = - i v \: \intop_{-\ell}^\ell \: d x \: \sum_{\lambda = 1}^3 \: \{\xi_{R,\lambda} (x) \partial_x \xi_{R,\lambda} (x) 
+ \zeta_{R,\lambda} (x) \partial_x \zeta_{R,\lambda} (x) \} \nonumber \\
&& +  i v \: \intop_{-\ell}^\ell \: d x \: \sum_{\lambda = 1}^3 \: \{\xi_{L,\lambda} (x) \partial_x \xi_{L,\lambda} (x) 
+ \zeta_{L,\lambda} (x) \partial_x \zeta_{L,\lambda} (x) \} 
\:\: .\nonumber \\
\label{ap.c.11}
\end{eqnarray}
\noindent
AT the DFP the open boundary conditions on the $\psi_\lambda$ fields at $x=0$ in Eqs.(\ref{ap.c.3}) 
translate into analogous conditions on the real fermionic fields, given by 

\begin{eqnarray}
&& \xi_{R,\lambda} (0) + \xi_{L,\lambda} (0) = 0 \nonumber \\
&& \zeta_{R,\lambda}(0) + \zeta_{L,\lambda} (0) = 0 
\:\: .
\label{ap.x1}
\end{eqnarray}
\noindent
Given Eqs.(\ref{ap.x1}), we unfold the chiral real fermionic fields and define two triples 
of real fields $\{\xi_\lambda (x) , \zeta_\lambda (x) \}$, with  $-\ell \leq x \leq \ell$, given by 

\begin{eqnarray}
\xi_\lambda (x) &=& \theta (x) \xi_{R,\lambda} (x) - \theta (-x) \xi_{L,\lambda } (-x) \nonumber \\
\zeta_\lambda (x) &=& \theta (x) \zeta_{R,\lambda} (x) - \theta (-x) \zeta_{L,\lambda} (-x) 
\:\: . 
\label{ap.x2}
\end{eqnarray}
\noindent
The partition function at the DFP in the $\xi$-sector, ${\cal Z}_{\xi}$, can be computed 
following the standard procedure of 1+1-dimensional conformal field theories (CFT)  \cite{DiFrancesco1997}.  
As a result, one obtains

\beq
{\cal Z}_{\xi} = \{  q^{-\frac{1}{24}} \: \prod_{m=0}^\infty \{1+q^{\left(m+\frac{1}{2}\right)} \} \}^3 
\;\; ,
\label{ap.d.3}
\eneq
\noindent
with $q=\exp \left[-\frac{\beta 2\pi v}{2 \ell} \right]$.  Expanding the right-hand side of Eq.(\ref{ap.d.3}) around
$q=0$, one gets, for the disconnected junction

\beq
{\cal Z}_{\xi}  = q^\frac{1}{16} \{1+3q^\frac{1}{2} + 3 q + 4 q^\frac{3}{2} + \ldots \}
\:\: .
\label{ap.d.4}
\eneq
\noindent
From the standard CFT analysis of   Eq.(\ref{ap.d.4}) 
we conclude that the ``1'' corresponds to the primary field given by the identity 
operator, the  ``$q^\frac{1}{2}$'' corresponds to the triple of primary fields realized by the 
real fermions of scaling dimensions $\frac{1}{2}$ \cite{DiFrancesco1997}. 
The ``$3q$'' accounts for the three, dimension-1, level-1 descendants of the identity, 
that is, the three currents ${\cal J}_\lambda (x)$ given by 

\beq
{\cal J}_\lambda (x) = -i\xi_{\lambda+1}(x) \xi_{\lambda+2} (x)
\;\;,
\label{ap.d.5}
\eneq
\noindent
which close the level-2, $su(2)$ affine algebra 

\beq
[{\cal J}_\lambda (x), {\cal J}_{\lambda'} (x')] = i \epsilon_{\lambda,\lambda',\lambda"} \delta (x-x') {\cal J}_{\lambda"} (x') 
-\frac{i}{2\pi} \delta^{'} (x-x')
\:\:,
\label{ap.d.6}
\eneq
\noindent
with $\epsilon_{\lambda,\lambda',\lambda''}$ being the Levi-Civita symbol. A similar construction 
holds for the $\zeta$-sector of the theory, as well. As a result, on turning on $J_1$ and $J_2$, 
we may rewrite the boundary Kondo Hamiltonian $H_{K,f}$  as 

\begin{eqnarray}
H_{K,f} &=&     \bar{J}_1   \: \sum_{\lambda = 1}^3 [ -i \xi_\lambda (0) \xi_{\lambda+1} (0) ] 
[-i\eta_\lambda \eta_{\lambda +1} ]  \nonumber \\
&+&   \bar{J}_2   
\: \sum_{\lambda = 1}^3 [ - i \zeta_\lambda (0 ) \zeta_{\lambda + 1} (0 ) ] [ - i \eta_\lambda \eta_{\lambda + 1} ] 
\:\: .
\label{ap.c.12}
\end{eqnarray}
\noindent   
Similarly, we may express the spin- and the heat-current operators as  

\begin{eqnarray}
{\cal J}_{S,\lambda}^z (x,t) &=&  i v \{ \xi_\lambda (x,t) \zeta_\lambda (x,t) - \xi_\lambda (-x,t) \zeta_\lambda (-x,t) \} \nonumber \\
{\cal J}_{Q,\lambda} (x,t) &=& - i v^2 \sum_{a = \pm 1} \{ \xi_\lambda (ax,t) \partial_x \xi_\lambda (ax,t ) \nonumber \\
&+& \zeta_\lambda (ax,t ) \partial_x \zeta_\lambda (ax,t ) \} 
\:\: . 
\label{ap.c.13}
\end{eqnarray}
\noindent
In the case in which $J_2=0$, the model described by $H_{\xi} + H_{K,f}$, 
with 

\beq
H_\xi = -iv \: \sum_{\lambda=1}^3 \: \intop_{-\ell}^\ell \: d x \: \xi_\lambda (x) \partial_x \xi_\lambda (x) \: 
\;\; , 
\label{ap.cz1}
\eneq
\noindent
 corresponds to the realization of the two-channel Kondo model 
(2CK) originally proposed and discussed in Ref.\cite{Coleman1995}. Within such a formulation of the 2CK, it 
is possible to construct a simple, field-theoretical description of the system at the Kondo FP, by
adapting to our specific case a trick tipically employed within the  CFT approach to 
 the Kondo effect \cite{Affleck1995}.    Specifically, we note that  
 
\beq
H_{\xi} +H_{K,f} =  \pi v \sum_{\lambda=1}^3 \: \intop_{-\ell}^\ell \: dx \: \left\{ [{\cal J}_\lambda (x)]^2 +  \frac{\bar{J}_1}{ \pi v}
{\cal J}_\lambda (x) {\cal T}_\lambda \delta (x) \right\}
\:\: . 
\label{ap.d.9}
\eneq
\noindent
As a result, on defining the shifted current triplet ${\cal I}_\lambda (x)$ as

\beq
{\cal I}_\lambda (x) = {\cal J}_\lambda (x) + \frac{\bar{J}_1}{2 \pi v} \:  {\cal T}_\lambda \delta (x)
\:\:,
\label{ap.d.10}
\eneq
\noindent
we see that, for $\frac{\bar{J}_1}{2 \pi v} = 1$, apart for an unessential constant, Eq.(\ref{ap.d.9}) can be 
rewritten as

\beq
H_{\xi} +H_{K,f}  =  \pi v \sum_{\lambda=1}^3 \: \intop_{-\ell}^\ell \: dx \: \{ {\cal I}_\lambda (x) \}^2
\:\:,
\label{ap.d.11}
\eneq
\noindent
with the triple ${\cal I}_\lambda (x)$ closing exactly the same affine algebra as ${\cal J}_\lambda (x)$ (see
Eq.(\ref{ap.d.6})).  As a result, we conclude that the 2CK  FP is the same as 
the disconnected FP, except that the local topological spin 
has been reabsorbed in the $su(2)$ affine current. As there is no longer the local topological spin
operator available, the leading boundary operator at the 2CK  FP has to be separately constructed, 
as we outline  below.

In fact, moving from the DFP to the  
${\cal I}_1$ FP just corresponds to  a pertinent change in the boundary conditions of the real fermionic
fields at the junction. To recover it, we  rewrite $H_{K,f}$ without previously assuming any specific boundary
condition on the fermionic fields as $H_{\rm K,f} = H_{\rm K,f,1} + H_{\rm K,f,2}$, with 

\begin{eqnarray}\label{ap.d.x3}
H_{\rm K,f,1} &=&  -i J_1 \: \sum_{\lambda=1}^3 ( \bar{\xi}_{R,\lambda} (0)+\bar{\xi}_{L,\lambda}(0) )  \times \nonumber \\  
&& \qquad ( \bar{\xi}_{R,\lambda+1} (0) 
+ \bar{\xi}_{L,\lambda+1} (0) )[-i\eta_\lambda \eta_{\lambda +1} ]  \nonumber \\
H_{\rm K,f,2} &=&   -iJ_2 \: \sum_{\lambda = 1}^3  (\bar{\zeta}_{R,\lambda}(0) + \bar{\zeta}_{L,\lambda}(0) ) \times  \\
&&\qquad (\bar{\zeta}_{R,\lambda+1}(0) + \bar{\zeta}_{L,\lambda+1}(0))
[-i\eta_\lambda \eta_{\lambda +1} ] \nonumber
\:\:,
\end{eqnarray}
\noindent
and 

\begin{eqnarray}
\bar{\xi}_{R,\lambda} (x) &=& \cos (k_F) \xi_{R,\lambda}(x) - \sin (k_F) \zeta_{R,\lambda}(x) \nonumber \\
\bar{\xi}_{L,\lambda} (x) &=& \cos (k_F ) \xi_{L,\lambda}(x) + \sin (k_F) \zeta_{L,\lambda} (x) \nonumber \\
\bar{\zeta}_{R,\lambda}(x) &=& \sin (k_F) \xi_{R,\lambda}(x) + \sin (k_F ) \zeta_{R,\lambda} (x) \nonumber \\
\bar{\zeta}_{L,\lambda} (x) &=& - \sin (k_F ) \xi_{L,\lambda} (x) + \cos (k_F ) \zeta_{L,\lambda}(x) 
\:  . 
\label{ap.z1}
\end{eqnarray}
\noindent
According to Eqs.(\ref{ap.z1}), when expressing in terms of the ``rotated'' fields the bulk Hamiltonian, we obtain 

\begin{eqnarray}
&& H_0 =\label{ap.z2}\\
&& - i v \sum_{\lambda = 1}^3 \: \int_0^\ell \: d x \: \{\bar{\xi}_{R,\lambda} (x) \partial_x \bar{\xi}_{R,\lambda}(x) + \bar{\zeta}_{R,\lambda}(x) 
\partial_x \bar{\zeta}_{R,\lambda}(x) \} \nonumber \\
&& + i v  \sum_{\lambda = 1}^3 \: \int_0^\ell \: d x \: \{\bar{\xi}_{L,\lambda} (x) \partial_x \bar{\xi}_{L,\lambda}(x) + \bar{\zeta}_{L,\lambda}(x) 
\partial_x \bar{\zeta}_{L,\lambda}(x) \}  \;\; . 
 \nonumber \end{eqnarray}
\noindent
From our previous discussion of the ${\cal I}_1$ FP we have realized  that it is characterized by the possibility of fully
reabsorbing the Kondo interaction in a pertinent redefinition of the $su(2)$ affine currents. From Eqs.(\ref{ap.d.x3},\ref{ap.z1},\ref{ap.z2}) 
we see that  the same result is recovered by imposing pertinent boundary conditions, at $x=0$, over the rotated 
real fermionic fields, that is 

\begin{eqnarray}
&& \bar{\xi}_{R,\lambda} (0) + \bar{\xi}_{L,\lambda} (0) = 0 \nonumber \\
&& \bar{\zeta}_{R,\lambda}(0) + \bar{\zeta}_{L,\lambda} (0) = 0 
\:\: . 
\label{ap.z3}
\end{eqnarray}
\noindent
The conditions in Eqs.(\ref{ap.z3}) are analogous to the ones characterizing the DFP. However, as a consequence 
of the nontrivial transformations in Eqs.(\ref{ap.z1}), they come along with different results for the spin- and heat-currents
at the ${\cal I}_1$ FP, as the corresponding operators are now expressed, in terms of the rotated fields, as

\begin{eqnarray}
{\cal J}^z_{S,\lambda} (x) &=& iv \: \sum_{a=\pm} \: \bar{\xi}_\lambda (ax) \bar{\zeta}_\lambda (ax) \nonumber \\
{\cal J}_{Q,\lambda}(x) &=& -iv^2 \: \sum_{a=\pm 1} \: a \times \nonumber \\
&& \{\bar{\xi}_\lambda (ax) \partial_x \bar{\xi}_\lambda (ax) 
+  \bar{\zeta}_\lambda (ax) \partial_x \bar{\zeta}_\lambda (ax) \}
\;\; , 
\label{strois.16}
\end{eqnarray}
\noindent
with the unfolded fields defined as 

\begin{eqnarray}
\bar{\xi}_\lambda (x) &=& \theta (x) \bar{\xi}_{R,\lambda}(x) - \theta (-x) \bar{\xi}_{L,\lambda}(-x) \nonumber \\
\bar{\zeta}_\lambda (x) &=& \theta (x) \bar{\zeta}_{R,\lambda}(x) - \theta (-x) \bar{\zeta}_{L,\lambda} (-x) 
\:\: . 
\label{ap.z4}
\end{eqnarray}
\noindent
In the main text we discuss the consequences of the modified formulas for the current operators for  
the transport properties at the junction. 

To conclude this Appendix, we now construct the leading boundary perturbation  of the system at the ${\cal I}_1$ FP, $H_{{\cal I}_1}$. 
 To construct $H_{{\cal I}_1}$, we move back to the lattice description of the junction. Accordingly, 
we write  the bulk Hamiltonian as:

\beq
H_{\rm Lat}  = -2iJ \: \sum_{\lambda = 1}^3 \: \sum_{j=-\ell}^{\ell -1} \{ 
\bar{\xi}_{j,\lambda} \bar{\xi}_{j+1,\lambda} + \bar{\zeta}_{j,\lambda} \bar{\zeta}_{j+1,\lambda} \}  
\:\: , 
\label{laf.zy2}
\eneq
\noindent
and $H_{K,f,1}$ as:

\begin{eqnarray}
H_{K,f,1} &=&   J_1 \: \sum_{\lambda=1}^3 \: [-i(\bar{\xi}_{1,\lambda} + \bar{\xi}_{-1,\lambda}) (\bar{\xi}_{1,\lambda+1} + \bar{\xi}_{-1,\lambda+1})] 
\times \nonumber \\
&& [-i \eta_\lambda \eta_{\lambda +1} ] 
\:\: . 
\label{laf.zz1}
\end{eqnarray}
\noindent
To minimize  $H_{K,f,1}$ in the large-$J_1$ limit, we introduce the  Dirac fermionic operators 

\beq
d_\lambda = \frac{1}{\sqrt{2}}  \left\{ \frac{\bar{\xi}_{1,\lambda} + \bar{\xi}_{-1,\lambda}}{\sqrt{2}} + i \eta_\lambda \right\} 
\;\; , 
\label{ap.c.14}
\eneq
\noindent
so that we may rewrite $H_{K,f,1}$ as 

\beq
H_{K,f,1} = - 2 J_1 \: \left\{ \sum_{\lambda = 1}^3 \left[ d_\lambda^\dagger d_\lambda - \frac{1}{2} \right] \right\} ^2 
+ \frac{3 \bar{J}_1}{2} 
\:\: . 
\label{ap.c.15}
\eneq
\noindent
From Eq.(\ref{ap.c.15}) we see that the groundstate of $H_{K,f,1}$ is twofold degenerate. 
Indeed, denoting with $|n_1,n_2,n_3\rangle$ the state with 
occupation number of mode $d_\lambda$ equal to $n_\lambda$ ($\lambda=1,2,3$), both $|0,0,0\rangle$ and 
$|1,1,1\rangle$ have energy $- 3J_1$. At variance, the higher energy eigenvalue $J_1$ is 
sixfold degenerate, as the corresponding eigenspace is spanned by all the states with one of the $n_\lambda$'s equal to 0
and the other two equal to 1, or vice versa. We now build the leading boundary perturbation at the (so far) ``putative'' 
2CK  FP, in which we have set $J_2=0$ and $J_1$ very large. To do so, we 
 connect the central triangle to the leads by singling out of the lattice bulk Hamiltonian the 
 term  containing the $\bar{\xi}_{1,\lambda}$ and the $\bar{\xi}_{-1,\lambda}$ operators. 
This is given by

\beq
H_t = -\frac{i \tau}{2} \: \sum_{\lambda=1}^3 \{\bar{\xi}_{1,\lambda} \bar{\xi}_{2,\lambda}  + \bar{\xi}_{-2,\lambda} \bar{\xi}_{-1,\lambda}  \} 
\;\; .
\label{strois.4}
\eneq
\noindent
Eventually we set $\tau=J$ but,  for the time being, we  assume that $\tau$ is a parameter unrelated to $J$.  
 Taking into account the 
 residual boundary operators at finite $J_2$ and $\tau$,  going through a systematic
Schrieffer-Wolff procedure involving boundary couplings and operators and resorting to 
the continuum fields, on  implementing the DEBC method of Ref.\cite{Oshikawa2006}, 
we obtain that the leading boundary perturbation is given by

\begin{eqnarray}
\hat{H}_{{\cal I}_1}  &=&  - {\cal A} [-i \prod_{\lambda = 1}^3 \bar{\xi}_\lambda (0) ]   {\cal Q} \nonumber \\
&+& i {\cal B} \sum_{\lambda = 1}^3 \{ \bar{\xi}_\lambda (0) \bar{\zeta}_{\lambda +1} (0) \bar{\zeta}_{\lambda +2}(0) ] \}
{\cal Q}  \nonumber \\
&+& {\cal C} \sum_{\lambda = 1}^3 \{ \bar{\xi}_\lambda (0) \bar{\xi}_{\lambda +1} (0) \bar{\zeta}_\lambda (0) 
\bar{\zeta}_{\lambda +1} (0) \} 
\;\; ,
\label{strois.17}
\end{eqnarray}
\noindent
 with ${\cal Q}$ being a real fermionic zero-mode operators connecting the degenerate minima of 
 $H_{K,f,1}$ \cite{Tsvelik2013,Giuliano2013}, and 

\begin{eqnarray}
{\cal A} &=& \frac{3\tau^3}{16 J_1^2} \nonumber \\
{\cal B} &=& \frac{\tau J_2^2}{8 J_1^2} \nonumber \\
{\cal C} &=& \frac{\tau^2 J_2}{\bar{J}_1^2}
\:\: . 
\label{srois.14}
\end{eqnarray}
\noindent
  To check the stability of ${\cal I}_1$, we employ the procedure of 
\cite{Chamon2003,Oshikawa2006,Affleck2013,Giuliano2019,Kane2020}, by noting how 
a straightforward counting of the scaling dimension of the operators entering the right-hand side
of Eq.(\ref{strois.17}) yields the values $d_{\cal A} = d_{\cal B} = \frac{3}{2}$, $d_{\cal C} = 2$. 
All the scaling dimensions are larger than the critical dimension 1 and, therefore, $\hat{H}_{{\cal I}_1} $ is 
an {\it irrelevant}   operator and the FP ${\cal I}_1$ (as well as ${\cal I}_2$) is, therefore, stable.
Yet, as we discuss in the main text, the residual interaction in 
Eq.(\ref{strois.17}) is responsible for a nonzero heat current pattern in the vicinity of the ${\cal I}_1$ and 
of the ${\cal I}_2$ FPs. Retaining only the less irrelevant contributions to $\hat{H}_{{\cal I}_1}$, 
we eventually recover the operator   $H_{{\cal I}_1}$   in Eq.(\ref{strois.17x}) of the main text.

\section{RG trajectory for $J_1 = J_2$}
\label{ren_sym}

In this section we discuss the symmetric case  $J_1=J_2 \equiv J$. In this case, we use   the   Hamiltonian $H_{\rm FL}$ in 
Eq.(\ref{ap.c.2})  complemented with the boundary Kondo Hamiltonian 

\begin{eqnarray}
H_\Delta &=& \frac{ J}{2} \: \sum_{\lambda =1}^3 \{ [-i \eta_\lambda \eta_{\lambda +1} ] 
 \times  \label{boso.3} \\
 && [-i [ (\psi_{R,\lambda}^\dagger (0) + \psi_{L,\lambda}^\dagger (0) ) 
 (\psi_{R,\lambda +1}  (0) + \psi_{L,\lambda+1}  (0) )\nonumber \\
 &+&  (\psi_{R,\lambda} (0) + \psi_{L,\lambda} (0) ) 
 (\psi_{R,\lambda +1}^\dagger  (0) + \psi_{L,\lambda+1}^\dagger  (0) ) ] \}  \:\: .\nonumber 
 \end{eqnarray}
 \noindent
 To achieve the (large-$J$) Kondo limit, we now apply chiral bosonization to Eqs.(\ref{boso.3}). To do so, we set  \cite{Michaeli2017}

\begin{eqnarray}
\eta_\lambda \psi_{R,\lambda} (x) &=& e^{ i  \sqrt{4 \pi} \varphi_{R,\lambda} (x) } \nonumber \\
\eta_\lambda \psi_{L,\lambda} (x) &=& e^{ i \sqrt{4\pi} \varphi_{L,\lambda} (x) } \;\; , 
\label{boso.4}
\end{eqnarray}
\noindent
with $\varphi_{R,\lambda}, \varphi_{L,\lambda}$ respectively being chiral, right-handed and left-handed
bosonic fields (defined for $0\leq x \leq \ell$). In terms of the chiral, bosonic fields, the spin-
and the heat-current operators
in lead-$\lambda$ are given by \cite{Giuliano2022,Buccheri2022} 

\begin{eqnarray}
{\cal J}_{S,\lambda} (x) &=& \frac{v}{\pi} \{\partial_x \varphi_R (x) + \partial_x \varphi_L (x) \} \nonumber \\
{\cal J}_{Q,\lambda} (x) &=& \frac{v^2}{\pi^2}  \{ (\partial_x \varphi_{R,\lambda}(x))^2 - ( \partial_x \varphi_{L,\lambda} (x))^2 \} 
\: . 
\label{boso.5}
\end{eqnarray}
\noindent
In order to go through the next steps, it is useful to switch to   the 
spin-plasmon fields $\phi_\lambda (x)$, together with their dual fields, $\theta_\lambda (x)$. These
are related to the chiral fields in Eqs.(\ref{boso.4}) via the relations 

\begin{eqnarray}
\phi_\lambda (x) &=& \varphi_{R,\lambda} (x) + \varphi_{L,\lambda} (x) \nonumber \\
\theta_\lambda (x) &=&  \varphi_{R,\lambda} (x) - \varphi_{L,\lambda} (x) 
\:\: . 
\label{boso.5x}
\end{eqnarray}
\noindent
As we discuss in the main text, the open boundary conditions imply that  the boundary interaction  reduces to

\begin{eqnarray}
 \tilde{H}_\Delta &=& - 4 \bar{J} \: \sum_{\lambda = 1}^3 \: \cos \left[ \sqrt{\pi}  (\phi_\lambda (0) - \phi_{\lambda +1} (0) ) \right]
\;\; . 
\label{boso.7}
\end{eqnarray}
\noindent
$\tilde{H}_\Delta$  in Eq.(\ref{boso.7}) corresponds to the $SO(3)$ TKM 
 \cite{Beri2012,Altland2014,Buccheri2012}, with Luttinger parameter $g=1$.  In this case, 
 $H_\Delta$ is marginally relevant \cite{Beri2012,Altland2013,Giuliano2019,Kane2020,Cardy1996,Giuliano2018}, 
 and drives the  system towards a large-$\bar{J}$  Kondo FP  \cite{Beri2012,Altland2014,Buccheri2012}. 
At the DFP the boundary conditions on the fermionic fields, $\psi_{R,\lambda}(0) + \psi_{L,\lambda}(0) = 0$, 
simply translate into Neumann-like boundary conditions for the bosonic fields $\phi_\lambda (x)$, that is, 
$\partial_x \phi_\lambda (0 ) = 0$, $\forall \lambda$. Now, as it is customary in analyzing junctions of quantum wires within bosonization framework \cite{Oshikawa2006}, we 
switch to the center-of-mass and to the relative chiral fields, $\Phi (x),\chi_{1,2} (x)$ (together with their 
dual ones), defined by 

\beq
\left[\begin{array}{c} \Phi (x) \\ \chi_1 (x ) \\ \chi_2 ( x ) \end{array} \right] 
= \left[ \begin{array}{ccc} \frac{1}{\sqrt{3}} & \frac{1}{\sqrt{3}} & \frac{1}{\sqrt{3}} \\ \frac{1}{\sqrt{2}} & - \frac{1}{\sqrt{2}} & 0 \\
\frac{1}{\sqrt{6}} & \frac{1}{\sqrt{6}} & - \frac{2}{\sqrt{6}} \end{array} \right] 
\left[ \begin{array}{c} \phi_1 ( x ) \\ \phi_2 ( x ) \\ \phi_3 ( x ) \end{array} \right] \;\;  , \label{boso.x12}
\eneq
\noindent
and by 

 \beq
\left[\begin{array}{c} \Theta (x) \\ \omega_1 (x ) \\ \omega_2 ( x ) \end{array} \right] 
= \left[ \begin{array}{ccc} \frac{1}{\sqrt{3}} & \frac{1}{\sqrt{3}} & \frac{1}{\sqrt{3}} \\ \frac{1}{\sqrt{2}} & - \frac{1}{\sqrt{2}} & 0 \\
\frac{1}{\sqrt{6}} & \frac{1}{\sqrt{6}} & - \frac{2}{\sqrt{6}} \end{array} \right] 
\left[ \begin{array}{c} \theta_1 ( x ) \\ \theta_2 ( x ) \\ \theta_3 ( x ) \end{array} \right] \:\: .
\label{boso.12}
\eneq
\noindent
In terms of the relative fields, we obtain \cite{Beri2012,Buccheri2022}

\beq
\tilde{H}_\Delta = 4\bar{J} \sum_{a=1}^3 \: \cos [ \sqrt{2 \pi} \hat{k}_a \cdot \vec{\chi} (0) ] 
\;\; , 
\label{boso.13}
\eneq
\noindent
with $\hat{k}_a = \left[ \cos \left(\frac{4\pi (a-1)}{3} \right) ,  \sin \left(\frac{4\pi (a-1)}{3} \right) \right]$, and 
$\vec{\chi} (0) = [\chi_1 (0), \chi_2 (0) ]$. 
The emergence of a boundary Hamiltonian such as 
$\tilde{H}_\Delta$ in Eq.(\ref{boso.13}) characterizes a number of different physical contexts, 
such as, for instance, the theory of the quantum Brownian motion on a triangular lattice \cite{Yi1995,Chamon2003}, or a junction
of three Bose liquids \cite{Tokuno2008}, or of three Josephson chains \cite{Giuliano2008,Giuliano2010}. At the 
DFP, the Neumann boundary conditions over the $\phi_\lambda$ fields translate into Neumann boundary 
conditions for the three fields $\Phi(x),\chi_{1,2}(x)$, at $x=0$. At variance, the 
FP that describes the system in the strongly coupled limit ($J \to \infty$) is recovered by minimizing 
the corresponding boundary interaction at the right-hand side of Eq.(\ref{boso.13}), that is, by requiring that

\beq
\sqrt{2 \pi} \hat{k}_ \cdot \vec{\chi} (0) = 2 \pi n_a \;\; , \; n_a \in {\cal Z}
\;\; , 
\label{boso.14}
\eneq
\noindent
forall $a=1,2,3$ (note that this naturally implies $n_1+n_2+n_3 = 0$). 
 As a result, we obtain that the Kondo FP is simply characterized by 
Dirichlet (rather than Neumann) boundary conditions at $x=0$ for the relative fields $\chi_{1,2} (x)$, while
the center-of-mass field $\Phi (x)$, being decoupled from the boundary interaction, still obeys Neumann boundary conditions. 

Clearly, all the discussion above applies in the case in which $J_1=J_2$ strictly. Consistently with the main 
phase diagram of our system, we expect that the TK FP is no longer stable, as soon as $J_1 \neq J_2$. 
To verify this point, let us set $\delta \bar{J} = \bar{J}_1-\bar{J}_2$ and let us consider the ``residual'' boundary Hamiltonian 
$\delta H_\Delta$ emerging 
at $\delta J \neq 0$. We obtain 

\begin{eqnarray}
&& \delta H_\Delta = \delta \bar{J} \sum_{\lambda =1}^3 \{[-i \eta_\lambda \eta_{\lambda +1} ]  \times 
\label{boso.17} \\
&&  [-i [ (\psi_{R,\lambda} (0) + \psi_{L,\lambda}  (0) ) 
 (\psi_{R,\lambda +1}  (0) + \psi_{L,\lambda+1}  (0) )\nonumber \\
 &+&  (\psi_{R,\lambda}^\dagger  (0) + \psi_{L,\lambda}^\dagger (0) ) 
 (\psi_{R,\lambda +1}^\dagger  (0) + \psi_{L,\lambda+1}^\dagger  (0) ) ]
\:\: . \nonumber
\end{eqnarray}
\noindent
Bosonizing the operator at the right-hand side of Eq.(\ref{boso.17}), we obtain 
\beq
\delta \tilde{H}_\Delta =- 4 \delta \bar{J} \: \sum_{\lambda = 1}^3 \: \cos \left[  \sqrt{\pi} 
(\phi_\lambda (0) + \phi_{\lambda +1} (0) ) \right] 
\;\; ,
\label{boso.19}
\eneq
\noindent
At the TK FP, taking into account the Dirichlet boundary conditions determined by the pinning of
$\chi_{1,2} (0)$, the only field effectively entering the right-hand side of Eq.(\ref{boso.19}) is 
the center of mass field $\Phi (x)$. That being stated, we find that
the  operator at the right-hand side of Eq.(\ref{boso.19}) becomes  (\ref{sf.13})

\bibliography{biblio.bib}
\end{document}